\documentclass[journal]{IEEEtran}  
\usepackage{enumerate}
\usepackage{url}
\usepackage{flushend}
\usepackage{multirow}
\usepackage{colortbl}
\usepackage{pifont}
\usepackage{setspace}
\usepackage{cite}
\usepackage{algorithm}
\usepackage{algorithmic}
\usepackage[subfigure]{graphfig}
\usepackage{amsthm,amsmath,amssymb}
\usepackage{booktabs}
\usepackage{subfigure}
\usepackage{graphicx}
\usepackage{mathrsfs}
\usepackage{makecell}
\usepackage{cleveref}

\begin{document}
\graphicspath{{simulationresults/}}

\title{Monitoring nonstationary processes based on recursive cointegration analysis and elastic weight consolidation}

\author{Jingxin Zhang, Donghua Zhou, ~\IEEEmembership{Fellow,~IEEE}, and Maoyin Chen, ~\IEEEmembership{Member,~IEEE}
\thanks{This work was supported by National Natural Science Foundation of China [grant numbers 62033008, 61751307, 61873143]. (Corresponding authors: Donghua Zhou; Maoyin Chen)}
\thanks{Jingxin Zhang is with the Department of Automation, Tsinghua University, Beijing 100084, China (e-mail: zjx18@mails.tsinghua.edu.cn). }
\thanks{Donghua Zhou is with College of Electrical Engineering and Automation, Shandong University of Science and Technology, Qingdao 266000, China and also with the Department of Automation, Tsinghua University, Beijing 100084, China (e-mail: zdh@mail.tsinghua.edu.cn).}
\thanks{Maoyin Chen is with the Department of Automation, Tsinghua University, Beijing 100001, China and also with School of Automation and Electrical Engineering, Linyi University, Linyi 276005, China (e-mail: mychen@tsinghua.edu.cn).}
\thanks{This paper has been submitted to IEEE Transaction on Cybernetics for potential publication.}
}

\maketitle
\IEEEpeerreviewmaketitle

\begin{abstract}

This paper considers the problem of nonstationary process monitoring under frequently varying operating conditions.
Traditional approaches generally misidentify the normal dynamic deviations as faults and thus lead to high false alarms. Besides, they generally consider single relatively steady operating condition and suffer from the catastrophic forgetting issue when learning successive operating conditions.
In this paper, recursive cointegration analysis (RCA) is first proposed to distinguish the real faults from normal systems changes,  where the model is updated once a new normal sample arrives and can adapt to slow change of cointegration relationship.
Based on the long-term equilibrium information extracted by RCA, the remaining short-term dynamic information is monitored by recursive principal component analysis (RPCA). Thus a comprehensive monitoring framework is built. When the system enters a new operating condition, the RCA-RPCA model is rebuilt to deal with the new condition. Meanwhile, elastic weight consolidation (EWC) is employed to settle the `catastrophic forgetting' issue inherent in RPCA, where significant information of influential parameters is enhanced to avoid the abrupt performance degradation for similar modes.
The effectiveness of the proposed method is illustrated by a practical industrial system.
\end{abstract}



\section{Background}

Process monitoring is increasingly significant and essential to guarantee the process safety \cite{khediri2011variable,baek2017empirical,yin2017fault,teixeira2014distributed,zhang2019process}. Approaches for stationary processes have been intensively investigated and considerable achievements have been obtained \cite{zhang2019an,rato2016a,yu2020Whole,yin2019decentralized,MinWang2020CEP}. However, process data are generally nonstationary due to varying load, changes of raw materials, aging of equipments and product grade transitions, etc \cite{shang2018recursive,yin2016an}.  This phenomenon is ubiquitous in industrial systems, for instance, the power stations, oil explorations, chemical processes, etc. It is urgent and challenging to investigate the monitoring techniques for nonstationary processes under various potential operating conditions \cite{tan2020nonstationary}. 

Recently, several methods have been developed for nonstationary process monitoring. Canonical variate analysis extracts dynamic latent information by state space formulations, which aims to reduce the order of dynamics and is generally applied to linear systems \cite{pilario2018canonical}. Dynamic latent variable models (DLVMs) extract dynamic and static latent components simultaneously, which extract the most predictable information first \cite{dong2020efficient}. The switching autoregressive DLVM was proposed for multimode processes in the probabilistic framework \cite{Zhou2018Multimode} and it requires that the model covers all operating modes. These methods aforementioned fail to distinguish the real faults from normal deviations under varying operating conditions, thus delivering high false alarm rates.  To settle this issue, slow feature analysis (SFA) was proposed to identify the real fault from the operating point deviation,  by separating dynamic information from steady state information \cite{shang2015concurrent}. SFA requires that the system operates in a particular steady condition \cite{zhang2019slow}, which is unsuitable for frequently varying operating conditions.

Cointegration analysis (CA) is an effective method to deal with nonstationary data \cite{Granger1987Co,johansen1988statistical} and able to distinguish the real faults from normal dynamic changes under various operating conditions.  It is based on the general consesus that the long-term equilibrium relationship, i.e., cointegration relationship, exists in physical and chemical processes because the nonstationary variables are correlated to each other and governed by specifical laws \cite{chen2009cointegration}.  When the cointegration relationship is broken, the system enters a new mode if the dynamic equilibrium relationship returns to normal.
Cointegration testing method was adopted primarily for nonstationary process monitoring in \cite{chen2009cointegration}.
Zhao \emph{et al}. intensively investigated CA and proposed several extensions of CA, including dynamic distributed strategy for large-scale processes \cite{zhao2019dynamic}, CA with SFA to establish a full-condition monitoring model \cite{zhao2018a}.

However, these CA-based methods assume that the cointegration relationship remains the same\cite{zhao2019dynamic,zhao2018a}, which is unrealistic in practical systems. Take the coal pulverizing system of power plant as an instance. The compositions and characteristics of one coal may change slowly  because they are influenced by environments and it is difficult to mix the coal quite evenly. Thus, the cointegration relationship would change accordingly.  Hansen \emph{et al}. presented a recursive form of cointegrated vector autoregressive models \cite{hansen1992recursive}, which could update the cointegration relationship to adapt to the new condition.  Another form of recursive CA was proposed to adapt to the slowly changing cointegration relationship and the model was updated based on a block of data \cite{yu2020recursive}. However, the monitoring consequences are affected by the length of data block and it is intractable to determine the optimal value. Only the dynamic information that reflected the control performance was extracted and the remaining information was neglected, thus causing insensitivity to detecting the faults that are orthogonal to cointegration space \cite{lin2019monitoringb}.

It is also a universal phenomenon in practical industrial systems that the cointegration relationship may change sharply and frequently. For instance, the type of coal changes in power plants frequently owing to the environmental requirements and economical benefits. The compositions and calorific value of different coals vary greatly. Assume that there are various variables and can be sorted into three blocks. One block of variables shares the similar trend and one block represents the critical manipulated variables, while the remaining variables are contained in another block.
Thus the cointegration relationship and some manipulated variables may be transited from one steady state to another.
It has been mentioned \cite{yu2020recursive} that it is necessary to establish a new CA model from scratch, to quickly adjust to the new cointegration relationship based on the newly collected data.  But the recursive CA failed to provide excellent performance \cite{yu2020recursive} because it requires abundant data.


Aimed at the issues mentioned above, this paper investigates the general nonstationary process monitoring, where the cointegration relationship and the manipulated variables would change from one steady state to another frequently. First, in order to distinguish the real faults from normal dynamic deviations, a novel version of recursive cointegration analysis (RCA) is proposed to track the long-term equilibrium relationship, where the CA model is updated once a new sample arrives. Based on RCA, we introduce recursive principal component analysis (RPCA) to deal with the remaining information, thus establishing a comprehensive monitoring framework. For convenience,  RCA with RPCA is denoted as RCA-RPCA.

When the cointegration relationship is recognized as broken by RCA and the dynamic equilibrium relationship returns to normal quickly, the system enters a new operating mode.  We need to retrain the RCA and RPCA models from scratch based on the new data. Here, we employ elastic weight consolidation (EWC) to settle the `catastrophic forgetting' issue of RPCA \cite{KirkpatrickOvercoming}, where the significant information that is influential in previous modes is preserved to avoid the dramatic performance degradation for similar operating modes.  For convenient description, the proposed RCA-RPCA with EWC is referred to as RCA-RPCA-EWC.
In addition, test statistics are established based on the prior knowledge and CA theory, which is more sensitive to normal change than recursive CA \cite{yu2020recursive}.

The rest of this paper is organized below. Section \ref{section:2} introduces the problem and reviews the basic theory of CA.
Section \ref{section:recursiveCA} presents the detailed procedure of the proposed RCA and summarizes the monitoring algorithm based on RCA-RPCA. Then, the proposed RCA-RPCA is extended to multimode processes in Section \ref{sec:pca-ewc}, where EWC is employed to overcome the `catastrophic forgetting' issue of RPCA when a new mode appears. Section \ref{section:monitoringmodel} summarizes the general procedure for nonstationary process monitoring, analyzes the computational complexity and compares with the state-of-the-art approaches. The effectiveness is illustrated by a practical industrial system in Section \ref{section:casestudy}. The concluding remark is presented in Section \ref{section:conclusion}.

\section{Problem formulation and preliminary}\label{section:2}

\subsection{Problem statement}

 Since there are various variables of multiple trends in practical applications, how to deal with the variables appropriately affects the monitoring performance severely. Take the practical coal pulverizing system of power plant as an instance.

\begin{figure}[htbp]
    	\centering
    	\subfigure{\label{variable-1}}\addtocounter{subfigure}{-2}
    	\subfigure
    	{\subfigure[CA variables]{\includegraphics[width=0.23\textwidth]{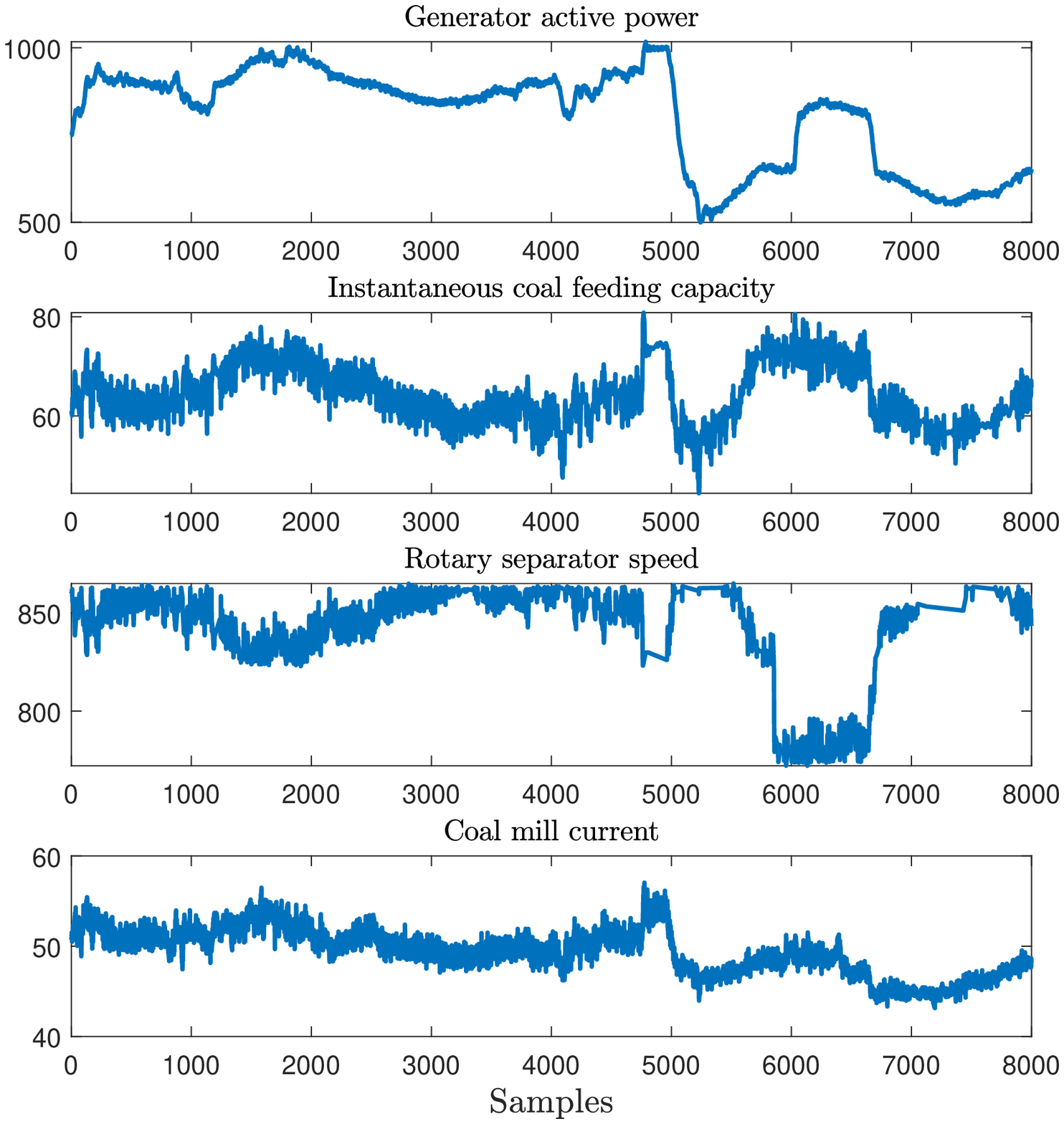}}}
    	\hspace{0.5mm}
    	\subfigure{\label{variable-2}}\addtocounter{subfigure}{-2}
    	\subfigure
    	{\subfigure[other variables]{\includegraphics[width=0.23\textwidth]{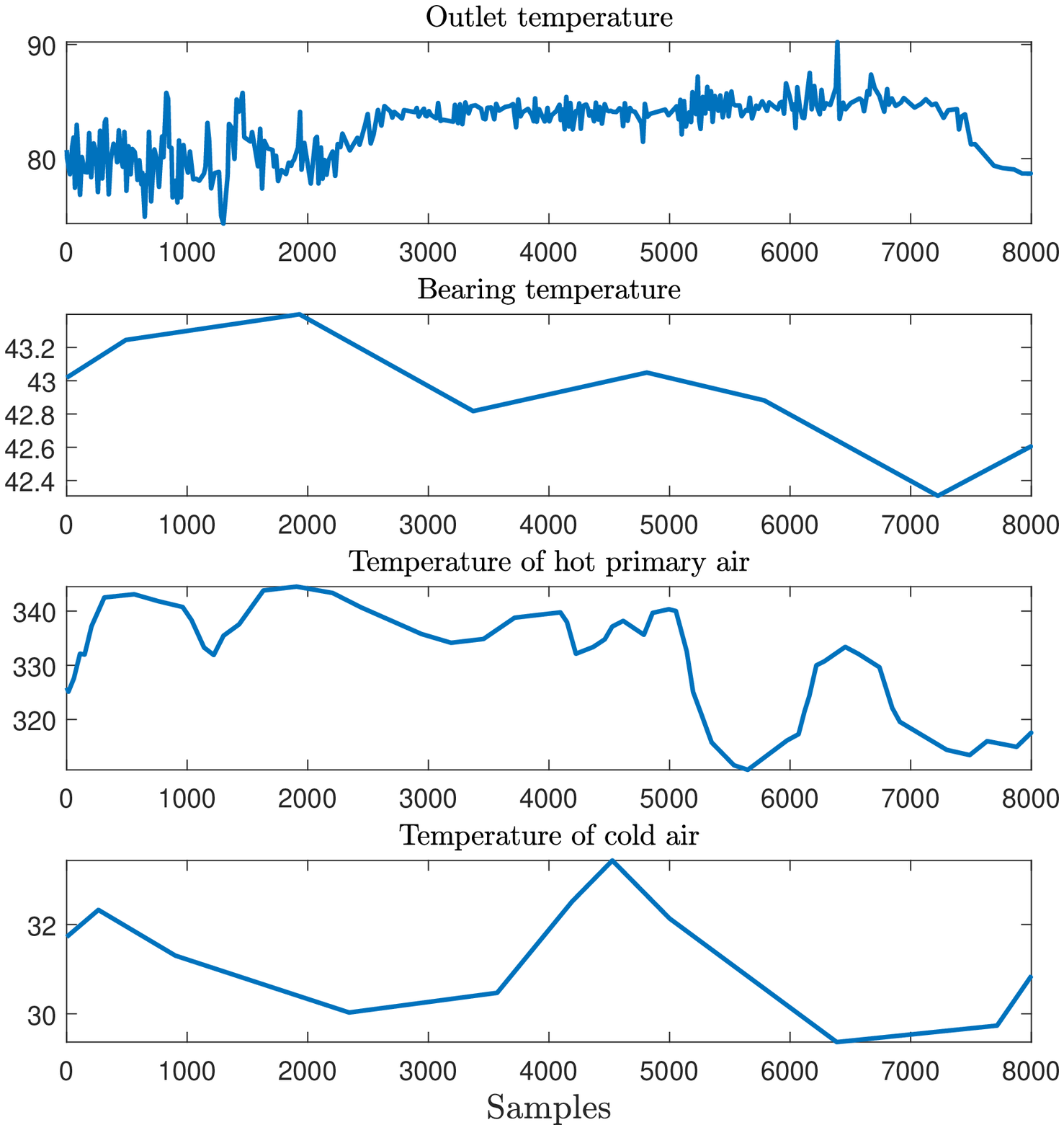}}}
        \caption{Practical data from the coal pulverizing system} \label{variable_description}
    \end{figure}

 The variables are affected by load and types of coal. Partial variables are described in Fig. \ref{variable_description}.
 The variables are decomposed into three blocks based on prior knowledge and augmented Dicky Fuller (ADF), where the final results rely on the prior knowledge and ADF test is the auxiliary to enhance universality.
 Variables in Fig. \ref{variable-1} are nonstationary and share the common trend, which are normally influenced by varying load.
For variables in Fig. \ref{variable-2}, the uppermost variable is regulated by controllers and the manipulated
variable is expected to vary  from one steady state to another one if the type of coal changes. The other three variables change slowly or irregularly.  The appropriate method needs to be investigated to deal with data of different characteristics, thus delivering an optimal monitoring performance. Note that this variable grouping makes it possible that this proposed method is sensitive to changes of manipulated variables and mode identification.

This paper studies the general case of sequential nonstationary process monitoring, where the stationary variables and the cointegration relationship change from one steady state to another.
RCA processes the data with common trend to extract long-term equilibrium information and RPCA is adopted to deal with other variables to extract short-term dynamics,  thus constructing a comprehensive  monitoring framework for nonstationary processes.  When the cointegrated relationship or stationary variables change, the system enters a new mode and  EWC is adopted to preserve the significant information of previous modes, thus delivering excellent performance for successive modes based on single model.



\subsection{Conventional CA algorithm}\label{traditionCA}


Given the nonstationary time series $\boldsymbol X^0=\left\{ \boldsymbol x^0_t \right\} _{t=1}^{N} $ with  $ \boldsymbol x^0_t \in \mathbb{R}^{m_1}$. 
The reference mean $ {\boldsymbol {\tilde  \mu_1}}$ and reference standard deviation $\tilde \sigma_1,\cdots,\tilde \sigma_{m_1}$ are calculated as
\begin{equation}\label{original_mean}
\boldsymbol {\tilde  \mu_1} = \frac{1}{N}(\boldsymbol X^0)^T \boldsymbol 1
\end{equation}
\begin{equation}\label{original_std}
 {{\tilde \sigma }_i} = \frac{1}{{N - 1}}\sum\limits_{t = 1}^N {{{\left( {x_{t,i}^0 - {{\tilde \mu }_i}} \right)}^2}} ,i \in \left\{ {1, \ldots ,{m_1}} \right\}
\end{equation}
where ${x_{t,i}^0}$ is the $i$th variable at $t$th sampling instant, $\boldsymbol 1$ is the vector of all ones with appropriate dimension. Thus, the original data $\boldsymbol X^0$ are normalized as
\begin{equation}
 \boldsymbol  X = \left( {{\boldsymbol  X^0} -  \boldsymbol 1{\boldsymbol {\tilde \mu }^T}} \right){ \boldsymbol{\tilde \Sigma }^{ - 1}}
\end{equation}
where $ \boldsymbol{\tilde \Sigma } = diag\left\{ \tilde \sigma_1,\cdots,\tilde \sigma_{m_1} \right\}$.

The vector error-correction (VEC) model is described as:
\begin{equation}\label{original_pro}
\varDelta \boldsymbol x_t=\sum_{i=1}^{p-1}{\boldsymbol \varOmega _i \varDelta \boldsymbol x_{t-i}+\boldsymbol \varGamma \boldsymbol x_{t-1}+ \boldsymbol \varepsilon_t}
\end{equation}
where ${\varDelta \boldsymbol x_t = \boldsymbol x_t- \boldsymbol x_{t-1}}$,
 $p$ is the order of VEC model  and determined by AIC. $\boldsymbol \varepsilon_t$ is the Gaussian white noise with ${\boldsymbol \varepsilon \sim N\left( \textbf{0},\boldsymbol \varXi \right)} $. ${\boldsymbol \varGamma =\boldsymbol \Upsilon  \boldsymbol B_f^T\in \mathbb{R}^{m_1\times m_1}}$, where ${\boldsymbol \Upsilon  \in \mathbb{R}^{m_1\times r}}$ and $ {\boldsymbol B_f \in \mathbb{R}^{m_1\times r}}$ are of full rank $r$. 
The columns in $\boldsymbol B_f$ are cointegration vectors. The objective of CA is to determine $\boldsymbol B_f$ to make the equilibrium errors $\boldsymbol X \boldsymbol B_f$ as stationary as possible.

Johansen \emph{et al}. proved that (\ref{original_pro}) could be settled by optimizing the likelihood function \cite{johansen1988statistical,hansen1999some}:
\begin{equation}
\begin{aligned}
& L\left(\boldsymbol \varOmega _1,\cdots ,\boldsymbol \varOmega _{p-1},\boldsymbol \Upsilon,\boldsymbol B_f,\boldsymbol \varXi \right)\\
&=-\frac{Nm_1}{2}\ln \left( 2 \right) -\frac{N}{2}\ln \left| \boldsymbol \varXi \right|-\frac{1}{2}\sum_{t=1}^N{\boldsymbol \varepsilon _{t}^{T}\boldsymbol \varXi ^{-1}\boldsymbol \varepsilon _t}
\end{aligned}
\end{equation}
The maximum likelihood estimation of cointegration vectors in $\boldsymbol B_f$ is acquired by eigenvalue decomposition (EVD) \cite{johansen1988statistical}  
\begin{equation}\label{CCA_eigen}
\left| \tilde{\lambda} \boldsymbol S_{11}- \boldsymbol S_{10} \boldsymbol S_{00}^{-1} \boldsymbol S_{01} \right|=0
\end{equation}
where $\boldsymbol S_{i,j}=\frac{1}{N-p} \boldsymbol E_{i}^{T} \boldsymbol E_j$, $\boldsymbol E_i $ ($ i =0,1$) is the prediction error and calculated by
\begin{equation}\label{e0}
\boldsymbol E_0=\varDelta \boldsymbol X_p-\varDelta \boldsymbol X^p \boldsymbol \varTheta
\end{equation}
\begin{equation}\label{e1}
\boldsymbol E_1=\boldsymbol X_p-\varDelta \boldsymbol X^p \boldsymbol \varPhi
\end{equation}
where $\varDelta \boldsymbol X_p \in \mathbb{R}^{\left( N-p \right) \times m_1}$  is the difference matrix, the vector $\varDelta \boldsymbol x_{p+1}= \boldsymbol x_{p+1}- \boldsymbol x_p$ is the temporal difference between two neighboring data points. $\boldsymbol X_p \in \mathbb{R}^{\left( N-p \right) \times m_1}$ originates from the observation matrix $\boldsymbol X$.  $\varDelta \boldsymbol X^p \in \mathbb{R}^{\left( N-p \right) \times pm_1}$ is the augmented matrix which contains $p$ lagged observations. The specific structures are described as
\begin{equation}\label{datamatrice1}
\boldsymbol X_p=\left[ \begin{array}{c}
\boldsymbol x_p\\
\boldsymbol x_{p+1}\\
\vdots\\
\boldsymbol x_{N-1}\\
\end{array} \right], \quad
\varDelta \boldsymbol X_p=\left[ \begin{array}{c}
\varDelta \boldsymbol x_{p+1}\\
\varDelta \boldsymbol x_{p+2}\\
\vdots\\
\varDelta \boldsymbol x_N\\
\end{array} \right]
\end{equation}
\begin{equation}\label{datamatrice2}
\varDelta \boldsymbol X^p=\left[ \begin{matrix}
\varDelta \boldsymbol x_1&			\cdots&		\varDelta \boldsymbol x_p\\
\vdots&				\ddots&		\vdots\\
\varDelta \boldsymbol x_{N-p}&				\cdots&		\varDelta \boldsymbol x_{N-1}\\
\end{matrix} \right]=\left[ \begin{array}{c}
\varDelta \boldsymbol x_{1}^{p}\\
\vdots\\
\varDelta \boldsymbol x_{N-p}^{p}\\
\end{array} \right]
\end{equation}

The coefficients $\boldsymbol \varTheta$ and $\boldsymbol \varPhi$ are obtained by ordinary least squares (OLS). Actually, (\ref{CCA_eigen}) can be reformulated as
\begin{equation}\label{generalized_original}
\boldsymbol A \boldsymbol w={\lambda} \boldsymbol B \boldsymbol w
\end{equation}
where $ \boldsymbol A=\left[ \begin{matrix}
\boldsymbol 0&		\boldsymbol S_{01}\\
\boldsymbol S_{10}&		\boldsymbol 0\\
\end{matrix} \right]
$, $\boldsymbol B=\left[ \begin{matrix}
\boldsymbol S_{00}&		\boldsymbol 0\\
\boldsymbol 0&		\boldsymbol S_{11}\\
\end{matrix} \right] $, the generalized eigenvalues are listed in the descending order.
$ \boldsymbol W = \left[ {{\boldsymbol w_1}, \cdots, {\boldsymbol w_r}} \right]\in \mathbb R^{2m_1\times r} $ contains the generalized principal eigenvectors  corresponding to $r$ largest eigenvalues and $r$ is determined by the trace test \cite{johansen1988statistical}. The cointegration matrix $\boldsymbol B_f$ and dynamic cointegration matrix  $\boldsymbol B_e$ are acquired from $\boldsymbol W$, namely, $\boldsymbol W = \left[ {{\boldsymbol B_e};{\boldsymbol B_f}} \right]$.
More information about CA can be found in \cite{Granger1987Co,johansen1988statistical}.

\section{The proposed RCA-RPCA for process monitoring}\label{section:recursiveCA}
In this section, we propose RCA to adapt to new cointegration relationship once a new sample arrives. The RCA issue is formulated into a recursive generalized EVD problem and settled by standard EVD. Besides, four test statistics are constructed according to prior knowledge and RCA-RPCA theory.

According to the prior knowledge and ADF test, this paper divides the variables into three blocks, one block represents the nonstationary variables with common trend, which are conducted by RCA and labeled as $\boldsymbol x_1$.  One block indicates the stationary variables that are sensitive to operating conditions and are denoted as $\boldsymbol x_2$. Generally, $\boldsymbol x_2$ is the critical manipulated variables and especially significant for industrial systems. The remaining block includes the variables independent of working conditions, which are expected to be stationary or change over the external environment and labeled as  $\boldsymbol x_3$. As a note, the variables are not necessarily divided into three blocks for any industrial system. It depends on the system characteristics and change regularities. However, the monitoring framework proposed in this paper also applies to this situation equally.

\subsection{Recursive cointegration analysis}\label{section:RCA}

We establish the initial CA model based on Section \ref{traditionCA}. If the cointegration relationship changes slowly, the collected data are preprocessed by fixed mean and standard deviation, as described in (\ref{original_mean}-\ref{original_std}).
The procedure of RCA is proposed below.

At $k+1$  instant, collect $\boldsymbol x_{k+1}^0$ and scale data as $\boldsymbol x_{k+1}$. The sample is divided into three blocks, namely, $\boldsymbol x_{k+1} = \left[{\begin{array}{*{20}{c}}
\boldsymbol x_{1,k+1}  &\boldsymbol x_{2,k+1} &\boldsymbol x_{3,k+1}
\end{array}}\right]$.  Only $\boldsymbol x_{1,k+1} $ is utilized for RCA.
Thus, the observations for RCA are $\boldsymbol  X_{1,k+1}=\left[ \begin{array}{c}
\boldsymbol X_{1,k}\\
\boldsymbol x_{1,k+1}\\
\end{array} \right]$.
Similar to (\ref{datamatrice1}-\ref{datamatrice2}), $\boldsymbol X_{p,k+1}$, $\varDelta \boldsymbol X_{p,k+1}$ and $\varDelta \boldsymbol X_{k+1}^p$ are generated from $\boldsymbol  X_{1,k+1}$.

The prediction errors are
\begin{equation}\label{E0k}
\boldsymbol E_{0,k+1}=\varDelta \boldsymbol X_{p,k+1}-\varDelta \boldsymbol X_{k+1}^p \boldsymbol \varTheta_{k+1}
\end{equation}
\begin{equation}\label{E1k}
\boldsymbol E_{1,k+1}=\boldsymbol X_{p,k+1}-\varDelta \boldsymbol X_{k+1}^p  \boldsymbol \varPhi_{k+1}
\end{equation}

According to recursive OLS,  $\boldsymbol \varTheta_{k+1}$ and $\boldsymbol \varPhi_{k+1}$ are determined by:
\begin{equation}\label{Thetak1}
\boldsymbol \varTheta _{k+1}=\boldsymbol \varTheta _k+ \boldsymbol R_{k+1}\left( \varDelta \boldsymbol x_{k+1}^{p} \right) ^T\left( \varDelta \boldsymbol x_{p,k+1}-\varDelta \boldsymbol x_{k+1}^{p} \boldsymbol \varTheta _k \right)
\end{equation}
\begin{equation}\label{Phik1}
\boldsymbol \varPhi _{k+1}=\boldsymbol \varPhi _k+\boldsymbol R_{k+1}\left( \varDelta \boldsymbol x_{k+1}^{p} \right) ^T\left( \boldsymbol x_{p,k+1}-\varDelta \boldsymbol x_{k+1}^{p}\boldsymbol \varPhi _k \right)\\
\end{equation}
where
\begin{equation}\label{Rk1}
\boldsymbol R_{k+1}=\boldsymbol R_k-\frac{\boldsymbol R_k\left( \varDelta \boldsymbol x_{k+1}^{p} \right) ^T\varDelta \boldsymbol x_{k+1}^{p}\boldsymbol R_k}{1+\varDelta \boldsymbol x_{k+1}^{p} \boldsymbol R_k\left( \varDelta \boldsymbol x_{k+1}^{p} \right) ^T}
\end{equation}
Bring (\ref{Thetak1},\ref{Rk1}) into (\ref{E0k}), then
\begin{equation}\label{E0K1}
\begin{aligned}
&\boldsymbol E_{0,k+1} \\
= &\varDelta \boldsymbol X_{p,k+1}-\varDelta \boldsymbol X_{k+1}^{p} \boldsymbol \varTheta _{k+1}\\
=&\left[ \begin{array}{c}
\varDelta \boldsymbol X_{p,k}\\
\varDelta \boldsymbol x_{p,k+1}\\
\end{array} \right] -\left[ \begin{array}{c}
\varDelta \boldsymbol X_{k}^{p}\\
\varDelta \boldsymbol x_{k+1}^{p}\\
\end{array} \right] \\
& \cdot  \left(\boldsymbol \varTheta _k+\frac{\boldsymbol R_k\left( \varDelta \boldsymbol x_{k+1}^{p} \right) ^T}{1+c_{k+1}}
\left( \varDelta \boldsymbol x_{p,k+1}-\varDelta \boldsymbol x_{k+1}^{p} \boldsymbol \varTheta _k \right) \right)
\\
= & \left[ \begin{array}{c}
\boldsymbol E_{0,k}-\frac{\varDelta \boldsymbol X_{k}^{p} \boldsymbol R_k\left( \varDelta \boldsymbol x_{k+1}^{p} \right) ^T}{1+c_{k+1}}\left( \varDelta \boldsymbol x_{p,k+1}-\varDelta \boldsymbol x_{k+1}^{p} \boldsymbol \varTheta _k \right)\\
\frac{1}{1+c_{k+1}}\left( \varDelta \boldsymbol x_{p,k+1}-\varDelta \boldsymbol x_{k+1}^{p} \boldsymbol \varTheta_k \right)\\
\end{array} \right]
\\
= & \left[ \begin{array}{c}
\boldsymbol E_{0,k}- \boldsymbol D_{k+1}\\
\boldsymbol d_{k+1}\\
\end{array} \right]
\end{aligned}
\end{equation}
where $\boldsymbol d_{k+1}=\frac{1}{1+c_{k+1}}\left( \varDelta \boldsymbol x_{p,k+1}-\varDelta \boldsymbol x_{k+1}^{p} \boldsymbol \varTheta _k \right)$,  $c_{k+1}=\varDelta \boldsymbol x_{k+1}^{p} \boldsymbol R_k\left( \varDelta \boldsymbol x_{k+1}^{p} \right) ^T$, $\boldsymbol D_{k+1}=\varDelta \boldsymbol X_{k}^{p} \boldsymbol R_k\left( \varDelta \boldsymbol x_{k+1}^{p} \right) ^T \boldsymbol d_{k+1}$.
Let $\boldsymbol J_k = \varDelta \boldsymbol X_{k}^{p} \boldsymbol R_k$, the recursion of $\boldsymbol J_k$ is
\begin{equation}
\begin{aligned}
\boldsymbol J_k&=\varDelta \boldsymbol X_{k}^{p} \boldsymbol R_k
\\
&=\left[ \begin{array}{c}
\varDelta \boldsymbol X_{k-1}^{p}\\
\varDelta \boldsymbol x_{k}^{p}\\
\end{array} \right] \left(\boldsymbol R_{k-1}-\frac{\boldsymbol R_{k-1}\left( \varDelta \boldsymbol x_{k}^{p} \right) ^T\varDelta \boldsymbol x_{k}^{p} \boldsymbol R_{k-1}}{1+c_k} \right)
\\
&=\left[ \begin{array}{c}
\boldsymbol J_{k-1}-\frac{\boldsymbol J_{k-1}\left( \varDelta \boldsymbol x_{k}^{p} \right) ^T\varDelta \boldsymbol x_{k}^{p} \boldsymbol R_{k-1}}{1+c_k}\\
\varDelta \boldsymbol x_{k}^{p} \boldsymbol R_{k-1}-\frac{\varDelta \boldsymbol x_{k}^{p}\boldsymbol R_{k-1}\left( \varDelta \boldsymbol x_{k}^{p} \right) ^T\varDelta \boldsymbol x_{k}^{p}\boldsymbol R_{k-1}}{1+c_k}\\
\end{array} \right] \\
&=\left[ \begin{array}{c}
\boldsymbol J_{k-1} \boldsymbol{\tilde{J}}_k\\
\varDelta \boldsymbol x_{k}^{p} \boldsymbol R_{k-1}\boldsymbol{\tilde{J}}_k\\
\end{array} \right]
\end{aligned}
\end{equation}
where $ \boldsymbol{\tilde{J}}_k =\boldsymbol I-\frac{\left( \varDelta \boldsymbol x_{k}^{p} \right) ^T\varDelta \boldsymbol x_{k}^{p} \boldsymbol R_{k-1}}{1+c_k} $, $\boldsymbol I$ is the identity matrix with appropriate dimension.
Thus, $\boldsymbol J_k$ and $ \boldsymbol D_{k+1}$ are calculated recursively. 
Similarly,
\begin{equation}\label{E1K1}
\begin{aligned}
&\boldsymbol E_{1,k+1} \\
= & \boldsymbol X_{p,k+1}-\varDelta \boldsymbol X_{k+1}^{p} \boldsymbol \varPhi _{k+1}
\\
=& \left[ \begin{array}{c}
\boldsymbol E_{1,k}-\frac{\varDelta \boldsymbol X_{k}^{p} \boldsymbol R_k\left( \varDelta \boldsymbol x_{k+1}^{p} \right) ^T}{1+c_{k+1}}\left( \boldsymbol x_{p,k+1}-\varDelta \boldsymbol x_{k+1}^{p} \boldsymbol \varPhi _k \right)\\
\frac{1}{1+c_{k+1}}\left( \boldsymbol x_{p,k+1}-\varDelta \boldsymbol x_{k+1}^{p} \boldsymbol \varPhi _k \right)\\
\end{array} \right]
\\
= & \left[ \begin{array}{c}
\boldsymbol E_{1,k}- \boldsymbol H_{k+1}\\
\boldsymbol h_{k+1}\\
\end{array} \right]
\end{aligned}
\end{equation}
where $\boldsymbol h_{k+1}=\frac{1}{1+c_{k+1}}\left(\boldsymbol x_{p,k+1}-\varDelta \boldsymbol x_{k+1}^{p} \boldsymbol \varPhi _k \right) $, $\boldsymbol H_{k+1}=\varDelta \boldsymbol X_{k}^{p} \boldsymbol R_k\left( \varDelta \boldsymbol x_{k+1}^{p} \right) ^T \boldsymbol h_{k+1} = \boldsymbol J_k \left( \varDelta \boldsymbol x_{k+1}^{p} \right) ^T \boldsymbol h_{k+1}$.

 Combining (\ref{E0K1}-\ref{E1K1}), $\boldsymbol A$ and $\boldsymbol B$ are calculated recursively as
 \begin{equation}\label{Cal_AK11}
 \begin{aligned}
 &\boldsymbol A_{k+1}\\
 =& \frac{1}{k+1}\left[ \begin{matrix}
 \boldsymbol 0&		\boldsymbol E_{0,k+1}^{T} \boldsymbol E_{1,k+1}\\
 \boldsymbol E_{1,k+1}^{T} \boldsymbol E_{0,k+1}&		\boldsymbol 0\\
 \end{matrix} \right] \\
 =&\frac{1}{k+1}\left(\left[ \begin{matrix}
 \boldsymbol 0&		\boldsymbol E_{0,k}^{T} \boldsymbol E_{1,k}\\
 \boldsymbol E_{1,k}^{T} \boldsymbol E_{0,k}&		\boldsymbol 0\\
 \end{matrix} \right] +\left[ \begin{matrix}
 \boldsymbol 0&		\varDelta {\boldsymbol A}_{1,k+1}\\
 \varDelta {\boldsymbol A}_{2,k+1}&		 \boldsymbol 0\\
 \end{matrix} \right]\right)
 \\
 =&\alpha_{k+1} \boldsymbol A_k +(1-\alpha_{k+1})\varDelta {\boldsymbol A}_{k+1}
 \end{aligned}
 \end{equation}
 where $\alpha_{k+1} = \frac{k}{k+1}$, $\varDelta {\boldsymbol A}_{1,k+1}=-\boldsymbol D_{k+1}^{T} \boldsymbol E_{1,k}-\boldsymbol E_{0,k}^{T} \boldsymbol H_{k+1}+ \boldsymbol D_{k+1}^{T} \boldsymbol H_{k+1}+\boldsymbol d_{k+1}^{T} \boldsymbol h_{k+1}$, $\varDelta {\boldsymbol A}_{2,k+1}=\varDelta {\boldsymbol A}_{1,k+1}^{T}$.
 \begin{equation}\label{Cal_BK11}
 \begin{aligned}
 &\boldsymbol B_{k+1}\\
 =&\frac{1}{k+1} \left[ \begin{matrix}
\boldsymbol E_{0,k+1}^{T} \boldsymbol E_{0,k+1}&		\boldsymbol 0\\
 \boldsymbol 0&		\boldsymbol E_{1,k+1}^{T} \boldsymbol E_{1,k+1}\\
 \end{matrix} \right] \\
 =&\frac{1}{k+1} \left(\left[ \begin{matrix}
\boldsymbol E_{0,k}^{T} \boldsymbol E_{0,k}&		 \boldsymbol 0\\
 \boldsymbol 0&		\boldsymbol E_{1,k}^{T} \boldsymbol E_{1,k}\\
 \end{matrix} \right] +\left[ \begin{matrix}
 \varDelta {\boldsymbol B}_{1,k+1}&		\boldsymbol 0\\
 \boldsymbol 0&		\varDelta {\boldsymbol B}_{2,k+1}\\
 \end{matrix} \right]\right)
 \\
 =&\alpha_{k+1} \boldsymbol B_k +(1-\alpha_{k+1})\varDelta {\boldsymbol B}_{k+1}
 \end{aligned}
 \end{equation}
 where $\varDelta {\boldsymbol B}_{1,k+1}=\boldsymbol D_{k+1}^{T}\boldsymbol D_{k+1} -  \boldsymbol D_{k+1}^{T} \boldsymbol E_{0,k}+ \boldsymbol d_{k+1}^{T} \boldsymbol d_{k+1}-\boldsymbol E_{0,k}^{T} \boldsymbol D_{k+1}$, $\varDelta {\boldsymbol B}_{2,k+1}= \boldsymbol H_{k+1}^{T} \boldsymbol H_{k+1}+ \boldsymbol h_{k+1}^{T} \boldsymbol h_{k+1}-\boldsymbol E_{1,k}^{T} \boldsymbol H_{k+1}- \boldsymbol H_{k+1}^{T} \boldsymbol E_{1,k}$.
Obviously, $ rank\left(\boldsymbol d_{k+1}^{T} \boldsymbol d_{k+1}- \boldsymbol E_{0,k}^{T}\boldsymbol D_{k+1} \right)=1 $, $ rank(\boldsymbol D_{k+1}^{T})=1 $. Thus, $ rank(\varDelta {\boldsymbol B}_{1,k+1}) \leqslant 2 $. Similarly, $ rank(\varDelta {\boldsymbol B}_{2,k+1}) \leqslant 2 $. $\varDelta {\boldsymbol A}$ and $\varDelta {\boldsymbol B}$ are also calculated recursively, as described in Appendix \ref{section:deltaAB}.

 The  proposed RCA is reformulated into settling the following generalized EVD problem:
 \begin{equation}\label{eigk00}
\boldsymbol A_{k+1} \boldsymbol W_{k+1}= \boldsymbol B_{k+1} \boldsymbol W_{k+1}\boldsymbol {\bar {\varLambda}}_{k+1}
 \end{equation}
where $\boldsymbol {\bar {\varLambda}}_{k+1}$ is the diagonal matrix and elements are generalized eigenvalues with descending order.

\subsection{Solution for numerical efficient recursive CA}		

In this paper, we convert a generalized EVD issue to a standard symmetric EVD problem.
As $\boldsymbol{B}_{\boldsymbol{k}+1}$ is symmetric and positive definite, let $ \boldsymbol K_{k+1}=\left( \boldsymbol B^{\frac{1}{2}}_{k+1}\right)^{-1} = \boldsymbol B^{-\frac{1}{2}}_{k+1}$, (\ref{eigk00}) can be reformulated as
\begin{equation}
 \boldsymbol{K}_{k+1}  \boldsymbol A_{k+1} \boldsymbol{K}^T_{k+1} \boldsymbol{\bar{W}}_{k+1} = \boldsymbol{\bar{W}}_{k+1}  \boldsymbol {\bar {\varLambda}}_{k+1}
\end{equation}
where $  \boldsymbol{K}_{k+1} $ is positive definite, $ \boldsymbol{\bar{W}}_{k+1} = \boldsymbol{K}^{-T}_{k+1}\boldsymbol W_{k+1}  $.  Computing $ \boldsymbol{K} $ directly may be ill-conditioning per update, thus it is essential to acquire the recursion of $ \boldsymbol{K} $ and avoid inverting a matrix repeatedly. The detailed derivation procedure is presented in Appendix \ref{section:recursion_K}.
To further reduce the computational burden, the recursion of $\boldsymbol K$ is obtained based on the rank of $\varDelta \boldsymbol B$.
The procedure for (\ref{eigk00}) is summarized in Algorithm \ref{ALG_power}.

\begin{algorithm}[!htbp]
	\caption{Solution based on Cholesky decomposition}\label{ALG_power}
	\begin{algorithmic}[1]
		\STATE  Calculate $\boldsymbol A_{k+1}$ by (\ref{Cal_AK11}) and  $\varDelta {\boldsymbol A}_{k+1}$ by Appendix \ref{section:deltaAB};
		
		\STATE  Calculate $\boldsymbol B_{k+1}$ by (\ref{Cal_BK11}) and $\varDelta {\boldsymbol B}_{k+1}$ by Appendix \ref{section:deltaAB};
		
		\STATE  Compute $\boldsymbol K_{k+1} = \boldsymbol B^{-\frac{1}{2}}_{k+1}$, as described in  Appendix \ref{section:recursion_K};

		\STATE Compute $ \boldsymbol C_{k+1} =\boldsymbol K_{k+1} \boldsymbol A_{k+1} \boldsymbol K^T_{k+1} $;
		
		\STATE Solve the eigenvalue problem of $ \boldsymbol C_{k+1} $ by symmetric QR algorithm, the eigenvectors and eigenvalues are denoted as $ \boldsymbol{\bar{W}}_{k+1} $ and $ \boldsymbol {\bar {\varLambda}}_{k+1}$, respectively; %
		
		\STATE Compute $ \boldsymbol W_{k+1} =  \boldsymbol K^T_{k+1} \boldsymbol{\bar{W}}_{k+1}$.

	\end{algorithmic}
\end{algorithm}

\subsection{Monitoring statistics}\label{monitoring_algorthm_single}

In this section, we construct the monitoring statistics to judge the operating conditions. The proposed RCA is utilized to extract the long-term equilibrium information and the short-term dynamic features are handled by RPCA. 
The key steps of RPCA have been elaborated in Appendix \ref{rank1-FOEP}.


At $k+1$ instant, a new sample is collected and preprocessed as $\boldsymbol x_{k+1} = \left[{\begin{array}{*{20}{c}}
\boldsymbol x_{1,k+1}  &\boldsymbol x_{2,k+1} &\boldsymbol x_{3,k+1}
\end{array}}\right]$. Let ${\hat {\boldsymbol x}_{1,k+1}} = \left[ {{\boldsymbol x_{1,k+1}}{\boldsymbol B_{f,k}}} \quad {{\boldsymbol x_{2,k+1}}} \right] $. The cointegration matrix $\boldsymbol B_{f,k}$  and dynamic cointegration matrix $\boldsymbol B_{e,k}$ are generated from generalized eigenvectors $\boldsymbol W_k$ in Algorithm \ref{ALG_power}.

$T^2_f$ is designed to judge whether the long-term static equilibrium relationship is still preserved.
\begin{equation}\label{t2f}
T^2_f = {\hat {\boldsymbol x}_{1,k+1}} {\hat {\boldsymbol x}_{1,k+1}}^T
\end{equation}

$T^2_e$ is designed to monitor the  long-term dynamic equilibrium relationship. 
\begin{equation}\label{t2e}
T^2_e  = {\boldsymbol e_{0,k+1}} \boldsymbol B_{e,k} \boldsymbol B_{e,k}^T{\boldsymbol e_{0,k+1}}^T
\end{equation}
where the prediction error ${\boldsymbol e_{0,k+1} } $  is the last sample of $\boldsymbol E_{0,k+1}$.

 Define $ \boldsymbol B_{f,k}^{\bot}=\boldsymbol I-\boldsymbol B_{f,k}\left(\boldsymbol B_{f,k}^{T}\boldsymbol B_{f,k} \right) ^{-1}\boldsymbol B_{f,k}^{T} $,  ${\hat {\boldsymbol x}_{2,k+1}} = \left[ {{\boldsymbol x_{1,k+1}}\boldsymbol B_{f,k}^{\bot}} \quad {{\boldsymbol x_{3,k+1}}} \right] $. The short-term dynamic information ${\hat {\boldsymbol x}_{2,k+1}}$  is monitored by RPCA and two statistics are calculated by
\begin{equation}\label{t2}
 T^2 = \hat{\boldsymbol x}_{2,k+1} \boldsymbol P_k   \boldsymbol \Lambda_k^{-1} \boldsymbol P_k^T  \hat{\boldsymbol x}^T_{2_k+1}
\end{equation}
\begin{equation}\label{spe}
SPE = \hat{\boldsymbol x}_{2,k+1} \left(\boldsymbol I -\boldsymbol P_k  \boldsymbol P_k^T \right) \hat{\boldsymbol x}^T_{2,k+1}
\end{equation}
where $\boldsymbol \Lambda_k$ and $\boldsymbol P_k$ represent  eigenvalues and eigenvectors of RPCA, which are updated by (\ref{RECUR1}-\ref{rpca_qv}) in Appendix \ref{rank1-FOEP}.

\section{Multimode process monitoring with EWC}\label{sec:pca-ewc}

In this section, we extend the nonstationary monitoring technique to multimode processes. Here, we define a mode where the stationary variables and the long-term static equilibrium fluctuate within acceptable range, which can be measured by $T^2_f$ and $T^2_e$ statistics. Actually, the data are still nonstationary in one mode.

When the system operates from one steady operating condition to another, the data distribution may change accordingly. Meanwhile, the cointegration relationship and the stationary variables may also vary dramatically. It has been illustrated that the recursive strategy of CA based on all collected data is unreasonable and may lead to high false alarms \cite{yu2020recursive}. RPCA also fails to track the rapid changes accurately. It is essential to build the proposed RCA-RPCA monitoring model from scratch. However, similar to most machine learning approaches \cite{ven2020brain,Zeng2019Continual,masse2018alleviating,zhang2020multimode}, RPCA suffers from the `catastrophic forgetting' issue and most information of the previous modes is overlapped when a new model is rebuilt. To settle this issue, EWC \cite{KirkpatrickOvercoming} is employed at the initial training phase of RPCA, where significant information from influential variables in previous modes is enhanced to avoid drastic changes. Thus, the proposed RCA-RPCA-EWC method can deliver outstanding monitoring performance when similar or the existing operating modes reappear.


Here, we introduce the procedure of RPCA with EWC (RPCA-EWC), which is similar to PCA with EWC in \cite{zhang2020multimode}.
Let $\boldsymbol  P^*_0$ denote the projection matrix for the previous operating mode.  When a new mode is detected by RCA,  the initial collected short-term dynamic data are denoted as $\boldsymbol X_2$.
The off-line training model of RPCA is built with EWC, thus the objective is designed as
 \begin{equation}\label{total_loss}
 \begin{aligned}
 	\mathcal{J}(\boldsymbol {P}) &= \mathcal{J}_{2}(\boldsymbol {P})+ \zeta  \mathcal{J}_{loss}(\boldsymbol {P},\boldsymbol  P^*_0)\\
 &=\|\boldsymbol X_2-\boldsymbol X_2 \boldsymbol P \boldsymbol P^T\|_F^2+ \|\boldsymbol  P - \boldsymbol  P^*_0 \|_{{\boldsymbol \Omega}}^2
 \end{aligned}
 \end{equation}
 where the hyperparameter $\zeta $  measures the importance of previous modes. The matrix ${\boldsymbol \Omega}$ is positive semidefinite, which is influenced by $\zeta $ and determined by \cite{zhang2020multimode,Husz2017On}. The constraint is $\boldsymbol  P^T \boldsymbol  P = \boldsymbol I$ with ${\boldsymbol {P} \in \mathbb{R}^{m_2 \times l}}$, $l$ is the number of principal components and determined by cumulative percent variance (CPV) approach.  $\mathcal{J}_2(\boldsymbol {P})$ is the loss function of RPCA for the current mode.
 $\mathcal{J}_{loss}(\boldsymbol P,\boldsymbol  P^*_0) $ is the loss function which measures the deviation of key parameters between two successive operating modes.

\begin{algorithm}[!tbp]
	\caption{Solution of RPCA-EWC}\label{EWC_solution}
	\begin{algorithmic}[1]
         \STATE  Let $\boldsymbol P_0=\boldsymbol P_0^*$ be the initial solution, error $\varepsilon$, set $i=0$;

         \STATE    Calculate $\boldsymbol Y_i= {\boldsymbol \Omega} \boldsymbol  P_0^*+ {\boldsymbol X_2^T}  {\boldsymbol X_2} \boldsymbol P_i$;

         \STATE Conduct singular vector decomposition on $\boldsymbol Y_i$, namely, ${\boldsymbol Y_i} = {\boldsymbol W_i}{\boldsymbol \Upsilon _i} \boldsymbol V_i^T$;

         \STATE   $\boldsymbol P_{i+1} = \boldsymbol W_i \boldsymbol I_{m,l} \boldsymbol V_i^T $;

         \STATE   Let $i = i+1$, go to 2 until $\lVert  \boldsymbol P_{i+1}-\boldsymbol P_i\rVert _F^2  < \varepsilon$.
 	\end{algorithmic}
\end{algorithm}

 The objective function (\ref{total_loss}) is actually the difference of convex (DC) functions programming problem \cite{Souza2015Global,Voorhis2003Difference}. DC programming includes linearizing the convex function and solving the convex function. The specific deviation process has been described in \cite{zhang2020multimode} and some key steps are listed in Appendix \ref{solution_RPCAEWC}. The solution is summarized in Algorithm \ref{EWC_solution}.
 Note that the matrix ${\boldsymbol \Omega}$ measures the importance of parameters and should be updated before a new mode appears. The calculation method can refer to \cite{Husz2017On,schwarz2018progress,zhang2020multimode}.

In summary, when  a new mode is judged by RCA, RCA model is rebuilt from scratch and the procedure is similar to Section \ref{section:recursiveCA}. RPCA-EWC is adopted at the initial training phase and then parameters are updated by (\ref{RECUR1}-\ref{rpca_qv}), thus avoiding abrupt degradation of monitoring performance when similar modes revisit.

\begin{algorithm}[!bp]
	\caption{Off-line training}\label{RCA-RPCA-EWC_training}
	\begin{algorithmic}[1]
      \STATE Collect the initial data $\boldsymbol X_{N_0}^0$, and set $k = N_0$;

		\STATE  According to prior knowledge and ADF test, divide $\boldsymbol X_k^0$ into three blocks, namely, $\boldsymbol X_{1,k}^0$, $\boldsymbol X_{2,k}^0$ and $\boldsymbol X_{3,k}^0$;

        \STATE Calculate the mean values and standard deviations of $\boldsymbol X_{1,k}^0$ and $\boldsymbol X_{2,k}^0$, i.e., $\hat {\boldsymbol \mu}_1$, $\hat{\boldsymbol \Sigma}_1$,  $\hat {\boldsymbol \mu}_2$, $\hat{\boldsymbol \Sigma}_2$. Scale data and denote as $\boldsymbol X_{1,k}$ and $\boldsymbol X_{2,k}$;

        \STATE Conduct CA on $\boldsymbol X_{1,k}$:

               a) Construct  $\boldsymbol X_{p,k}$, $\varDelta \boldsymbol X_{p,k}$ and $\varDelta \boldsymbol X_{k}^p$ by (\ref{datamatrice1}-\ref{datamatrice2});\\
               b) Calculate  coefficients of (\ref{e0}-\ref{e1}), labeled by $\boldsymbol \Theta_k$ and $\boldsymbol \Phi_k$;\\
               c) Calculate $\boldsymbol E_{0,k}$ and $\boldsymbol E_{1,k}$ in (\ref{e0}-\ref{e1}), and compute $ \boldsymbol A_k$ and $\boldsymbol B_k$;\\
               d) Solve (\ref{generalized_original}) and obtain $\boldsymbol B_{f,k}$, $\boldsymbol B_{e,k}$;

        \STATE Calculate $\boldsymbol B_{f,k}^{\bot} $ and construct ${\hat {\boldsymbol X}_{2,k}^0} = \left[ {{\boldsymbol X_{1,k}}\boldsymbol B_{f,k}^{\bot}} \quad {{\boldsymbol X_{3,k}^0}} \right] $.  Calculate the mean $\boldsymbol \mu_k$ and standard deviation $\boldsymbol \Sigma_k$, scale data and denote as $\hat {\boldsymbol X}_{2,k}$;

        \STATE Conduct PCA on ${\hat {\boldsymbol X}_{2,k}}$, and calculate $\boldsymbol P_{k}$ and $\boldsymbol \Lambda_{k}$;

        \STATE  Calculate test statistics by (\ref{t2f}-\ref{spe}) and the corresponding thresholds by KDE.
	\end{algorithmic}
\end{algorithm}

\section{Monitoring algorithm}\label{section:monitoringmodel}



State-of-the-art approaches explore the nonstationary processes for a single mode \cite{pilario2018canonical,lin2019monitoringb,zhao2019dynamic,zhao2018a}, where the stationary variables and the long-term static equilibrium fluctuate within a certain range. When the operating mode changes, the data distribution may vary accordingly and the original cointegration relationship is broken. This section introduces the general monitoring framework for multimode processes, which is also appropriate for a single mode.

Similar to Section \ref{section:recursiveCA}, the normal data are divided into three blocks.
At the training phase, the long-term equilibrium information is extracted by CA and PCA is utilized to monitor the remaining short-term dynamic information.
Four test statistics are calculated by (\ref{t2f}-\ref{spe}), where $T_f^2$ and  $T_e^2$ are employed to identify the operating status and $T^2$ and SPE are utilized to monitor the short-term dynamics. The corresponding thresholds are calculated by kernel density estimation (KDE) \cite{zhang2019an}. The off-line training procedure is summarized in Algorithm \ref{RCA-RPCA-EWC_training}.

For the practical industrial applications, when a new sample arrives, the operating status is judged and the monitoring model is updated if normal, as described in Algorithm \ref{RCA-RPCA-EWC_monitoring}. The thresholds are updated by KDE. Note that an occasional anomaly is regarded as noise or disturbance. The fault is detected if the anomaly lasts a short time.

The monitoring rule is summarized below:
\begin{enumerate}
	\item All test statistics are within their thresholds, it is regarded that the process operates normally in the same operating mode. The proposed RCA-RPCA is still employed to update the parameters; 
	
	\item  If $T^2_e$, $T^2$ and $SPE$ return to normal after $T^2_f$ is over its threshold, it indicates that the system enters a new operating state and then RCA-RPCA-EWC is adopted to monitor the system;
	
	\item If $T^2_f$ and $T^2_e$ are within their thresholds, while $T^2$ or $SPE$ is over its threshold, then a fault may occur and it is essential to check the operation of the systems;
	
	\item All test statistics exceed their thresholds, then the process is out of control. A real fault is  detected and the alarm is triggered.
\end{enumerate}

\begin{algorithm}[!tbp]
	\caption{Online monitoring}\label{RCA-RPCA-EWC_monitoring}
	\begin{algorithmic}[1]
        \STATE Collect $\boldsymbol x_{k+1}^0$, divide the sample into three blocks based on step 2 in Algorithm \ref{RCA-RPCA-EWC_training}, and scale data;

      \STATE Construct and scale $\hat{\boldsymbol x}_{1,k+1}$ and  $\hat{\boldsymbol x}_{2,k+1} $, calculate test statistics by (\ref{t2f}-\ref{spe});

        \STATE Judge the operating status: \\
        a) Normal, go to step 4; \\
        b) A new mode appears.  $n_0$ samples are collected, set $k=n_0$, go to step 2 in Algorithm \ref{RCA-RPCA-EWC_training}. Set $\boldsymbol X_2 = \hat{\boldsymbol X}_{2,k}$ and the step 6 is replaced by RPCA-EWC in Algorithm \ref{EWC_solution}; \\
        c) Potential fault or real fault occurs, thus the alarm is triggered;

        \STATE Conduct RCA based on the current CA model and $\boldsymbol x_{1,k+1}$:

               a) Construct $\boldsymbol x_{p,k+1}$, $\varDelta \boldsymbol x_{p,k+1}$ and $\varDelta \boldsymbol x_{k+1}^p$;\\
               b) Calculate  $\boldsymbol \Theta_{k+1}$ and $\boldsymbol \Phi_{k+1}$  by (\ref{Phik1}-\ref{Rk1});\\
               c) Calculate  $\boldsymbol E_{0,k+1}$ and $\boldsymbol E_{1,k+1}$ by (\ref{E0K1}-\ref{E1K1}), and compute $ \boldsymbol A_{k+1}$ and $\boldsymbol B_{k+1}$ by (\ref{Cal_AK11}-\ref{Cal_BK11});\\
               d) Solve  (\ref{eigk00}) by Algorithm \ref{ALG_power}, and obtain $\boldsymbol B_{f,k+1}$, $\boldsymbol B_{e,k+1}$;

        \STATE Construct  $\tilde {\boldsymbol x}_{2,k+1} = \left[ {{\boldsymbol x_{1,k+1}}\boldsymbol B_{f,k+1}^{\bot}} \quad {{\boldsymbol x_{3,k+1}}} \right]$, conduct RPCA:\\
               a) Calculate $\boldsymbol \mu_{k+1}$ and $\boldsymbol  \Sigma_{k+1}$ by (\ref{meank}-\ref{stdk});\\
               b) Calculate  $\boldsymbol P_{k+1}$ and $\boldsymbol \Lambda_{k+1}$ by (\ref{RECUR1}-\ref{rpca_qv});\\
               c) Select $l$ based on CPV;

        \STATE Set $k=k+1$ and return to step 1.
	\end{algorithmic}
\end{algorithm}

 \subsection{Computational complexity analysis}

 For online monitoring phase, the computational complexity contains the computation of RCA and RPCA at each step, and RPCA-EWC when the operating mode changes. The  RCA and RPCA algorithms occupy the most computational source and are considered in this paper.

 For RCA, the computation focuses on Algorithm \ref{ALG_power}. The complexity of $ \varDelta \boldsymbol A_{k+1} $ and $ \varDelta \boldsymbol B_{k+1} $ is $ O(m_1^2) $, as illustrated in Appendix \ref{section:deltaAB}. The complexity of $\boldsymbol K_{k+1}$ is $ O(m_1^3) $ in Appendix \ref{section:recursion_K}.
 Then, the calculation of $\boldsymbol C_{k+1}$ needs $8m_1^3$ flops.  The symmetric QR algorithm requires at most $32m_1^3$ flops theoretically because $\boldsymbol C_{k+1}$ is a block skew diagonal matrix.  The calculation of $\boldsymbol W_{k+1}$ in Algorithm \ref{ALG_power} requires $8m_1^3$ flops. In summary, the computational complexity of RCA is $O(m_1^3)$ per update.
  For RPCA, the complexity of $\boldsymbol P$ and $\boldsymbol \Lambda$  is $O(m_2^3+m_2^2)$. Obviously, ${m_1} < m$, ${m_2} \le  m$, $m$ is the dimension of collected data.
  That is, the computation per update will not grow as the number of samples $k$ increases.


\subsection{Comparison and Discussion}

We compare the recursive CA \cite{yu2020recursive} with the proposed RCA-RPCA-EWC method below:

  $\bullet$ Model tracking accuracy. The recursive CA model is updated  based on a block of data \cite{yu2020recursive}, and it is intractable to determine the data length to deliver the optimal performance. However, the proposed RCA model is updated once a new normal sample arrives.
  In the case that the cointegration relationship changes sharply and frequently, the recursive CA \cite{yu2020recursive} may fail to track the normal change, while the proposed approach can establish the inexact model based on just a few data and correct the model gradually.


  $\bullet$ Sensitivity of mode switching identification. The operating status is judged by $T_f^2$ and $T_e^2$. The construction of two statistics is only based on nonstationary data that reflect the control performance \cite{yu2020recursive}. The proposed statistics consider the prior knowledge and data simultaneously, which is more sensitive to mode switching.

  $\bullet$ Memory properties. The EWC technique is adopted to overcome the `catastrophic forgetting' issue of RPCA and significant information of previous operating conditions is enhanced to avoid dramatic changes of influential parameters.  The proposed RCA-RPCA can be updated accurately when previous or similar operating modes appear, thus delivering optimal monitoring performance.


  $\bullet$ Algorithm complexity.  The computational burden is highly related to the number of current collected samples at each update step \cite{yu2020recursive}. Although the models stop to update to reduce complexity and false alarm, it is hard to satisfy the criterion. For the proposed method, the computational cost is $O(m^3)$ per update, irrespective of the number of samples.

Here we make a further discussion about the variable decomposition. In this paper, the variables are divided into three blocks based on prior knowledge and data, as mentioned in Section III. Specifically, we first employ the theory of industrial systems to partition variables. Then we adopt ADF test and correlation analysis to verify and strengthen the rationality of variable grouping. Thus, the changes of stationary variables would not be covered by the normal variations of nonstationary variables.
 Note that it is not necessary that the variables are decomposed into three blocks in any industrial system. The number of blocks relies on the characteristics of systems and selected variables. However, the monitoring framework in Section III is also applied.  This variable grouping method is sensitive to critical manipulated variables and mode identification, which is  beneficial to enhance monitoring performance.



 \section{Case study}\label{section:casestudy}
This section adopts a practical industrial system to illustrate the effectiveness of the proposed method. Besides, we make a comparative analysis with the state-of-the-art methods to highlight the superiorities of the proposed method.

   \begin{figure}[!tbp]
 	\centering
 	\includegraphics[width=0.34\textwidth]{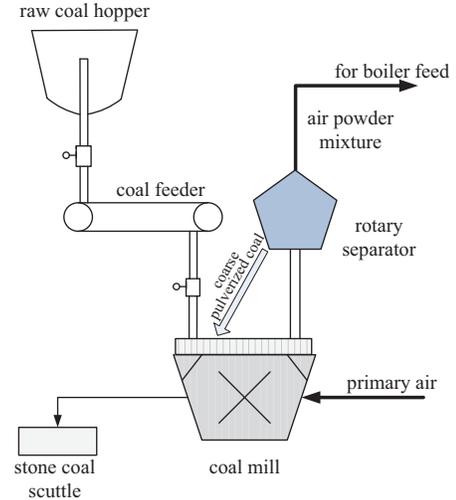}
 	\caption{Schematic diagram of the coal pulverizing system}
 	\label{fig_benchmark}
 \end{figure}

\begin{table*}[!htbp]
  	\begin{center}
  		\caption{Data information of the pulverizing system}\label{case_description}
  		\footnotesize
     \renewcommand{\baselinestretch}{1.2}
       \small
  		\begin{tabular}{l c c c c c c}
  			\hline
  			  Case  number     & Data original     &  \makecell{Training samples} & \makecell{Testing samples} & Fault time   & Fault cause\\
  			\hline
  Case 1 & Aomei-Aomeng-Aomei  & 2000  & 8800 & 6734 & \makecell{The opening of the regulating baffle \\of the primary air is abnormally large}\\
  Case 2 & Aomeng-Youhun      & 2000 & 15280 & 8731 & \makecell{Abnormality from cold primary air \\electric regulating baffle card} \\
  Case 3 & Fudong-Aomeng-Fudong & 2000 &13120 & 9010 &The  cooling fan motor trip \\
  			\hline
  		\end{tabular}
  	\end{center}
  \end{table*}

	\begin{table*}[!htbp]
 	\begin{center}   
 		\caption{Evaluation indexes of the case study}\label{Table1}
 		\renewcommand\arraystretch{1.3}
 		\small
 		\begin{tabular}{c c c c c c c c c c c c c } 
 			\hline
\multirow{2}{*}{Case number} & \multirow{2}{*}{Indexes} & \multicolumn{3}{c}{Recursive CA \cite{yu2020recursive}}  & \multicolumn{4}{c}{RCA-RPCA}  & \multicolumn{4}{c}{RCA-RPCA-EWC}    \\
   \cmidrule(r){3-5}  \cmidrule(r){6-9} \cmidrule(r){10-13}
        &     & \multicolumn{1}{c} {$T^2$}    & {$S^2$}   & {$D^2$}   & {$T_f^2$}   &$T_e^2$    &{$T^2$}  &{SPE}     &{$T_f^2$}   &$T_e^2$    &{$T^2$}  & {SPE}   \\
 \cline{1-13}
  \multirow{3}{*}{Case 1} & FDRs($\%$)  &59.31   & 0.19     &0        & 67.49 &99.85  & 59.75 & 66.81          & 75.91 & 99.85 &86.31  &69.04 \\
                           & FARs($\%$) & 0     & 0.16     &0         & 2.52 & 2.85   & 3.19  & 2.75           & 2.61  & 2.85  &4.72   &1.65 \\
                           & DD         & 841   & 847      &-      & 65   &3       & 832   & 686            & 128   & 3     & 283   & 630\\
  \cline{1-13}
  \multirow{3}{*}{Case 2} & FDRs($\%$) & 100    & 99.95    &33.88    & 12.34  & 99.98  &0  & 0              & 4.19  & 99.98  & 0     &93.86 \\
                           & FARs($\%$) & 20.01   &19.55   &7.92      & 3.95   &3.76   & 3.22 & 5.20        & 3.04  &3.56    &4.53   & 7.62 \\
                           & DD         &  0      & 0      &2         & 34     &1      & - & -        & 205   &1       & -  & 0\\

  \cline{1-13}
  \multirow{3}{*}{Case 3} & FDRs($\%$) & 86.21   & 0.05     &0         & 93.07 & 99.81 &84.41  & 89.42      & 95.26 & 99.46 &97.20 &86.72 \\
                           & FARs($\%$) & 0.71   & 0        &0          & 1.58  &3.65   & 1.42  & 1.15       & 1.43  & 3.51  &1.64 &1.40\\
                           & DD         & 0      & -        &-         & 43    &2     & 622   & 353         & 19    &6      & 113 & 544\\
  \cline{1-13}
 \end{tabular}
 \end{center}
 \end{table*}

 \subsection{Description of the pulverizing system}
 The 1000-MW ultra-supercritical thermal power plant is increasingly popular owing to economic benefits and environmental requirements. In this paper, we investigate one important unit of boiler, namely, the coal pulverizing system in Zhoushan Power Plant, Zhejiang Province, China \cite{zhang2020multimode}. It contains coal feeder, coal mill, rotary separator, raw coal hopper and stone coal scuttle, as depicted in Fig. \ref{fig_benchmark}. The coal pulverizing system grinds the raw coal into pulverized coal with desired coal fineness and optimal temperature. The operating conditions would change over the types of coal and varying unit load. For different types of coal, the cointegration relationship may change and the controlled variables may work at different stable points.

We choose 26 key variables and some typical variables have been depicted in Fig. \ref{variable_description} to illustrate the data characteristics.  Variables in Fig. \ref{variable-1} are relevant to the unit load, which are also nonstationary by ADF test and prior knowledge. Variables in Fig. \ref{variable-2} are little correlated with load. For instance, the air powder mixture temperature is required to be stationary and may be different for different types of raw coal. When the coal changes, the temperature would vary from one stable value to another one. The bearing temperatures are expected to remain at a stable level. The temperature of cold air is closely related to the external environment.
%
%

 We select three typical cases to illustrate the effectiveness of the proposed method, namely, abnormality from outlet temperature (Cases 1 and 2) and rotary separator (Case 3).  According to the historical records, these two types of faults occur frequently and affect the working safety.
 The sample interval is 20 seconds.  The data information is listed in Table \ref{case_description}. For each case, the process data come from two types of coal and the original cointegration relationship may be broken when the type of coal changes.

   \begin{figure*}[!htbp]
    	\centering
    	\subfigure{\label{fault1-1}}\addtocounter{subfigure}{-1}
    	{\subfigure[Recursive CA]{\includegraphics[width=0.33\textwidth]{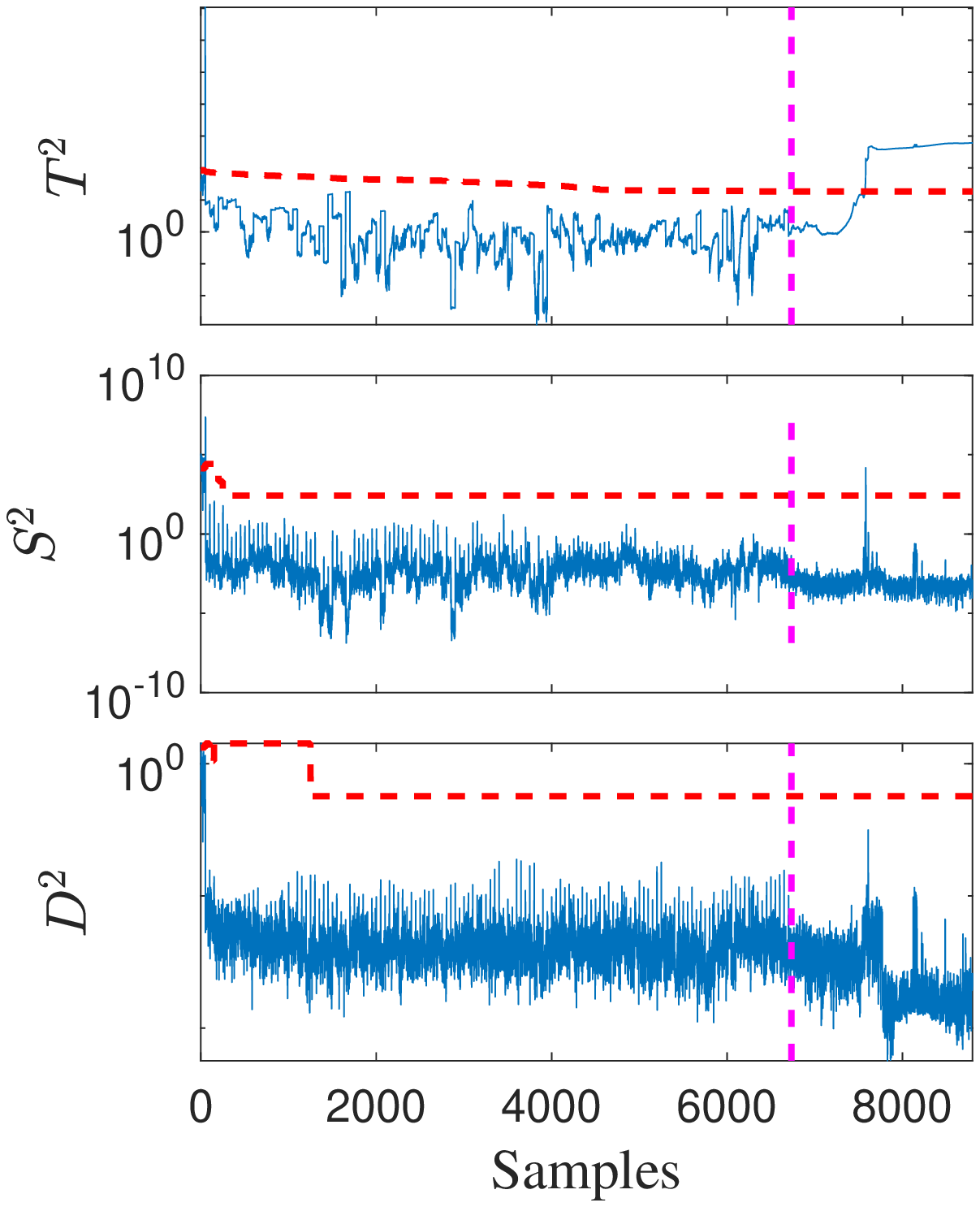}}}
    	\subfigure{\label{fault11-2}}\addtocounter{subfigure}{-1}
    	{\subfigure[RCA-RPCA]{\includegraphics[width=0.315\textwidth]{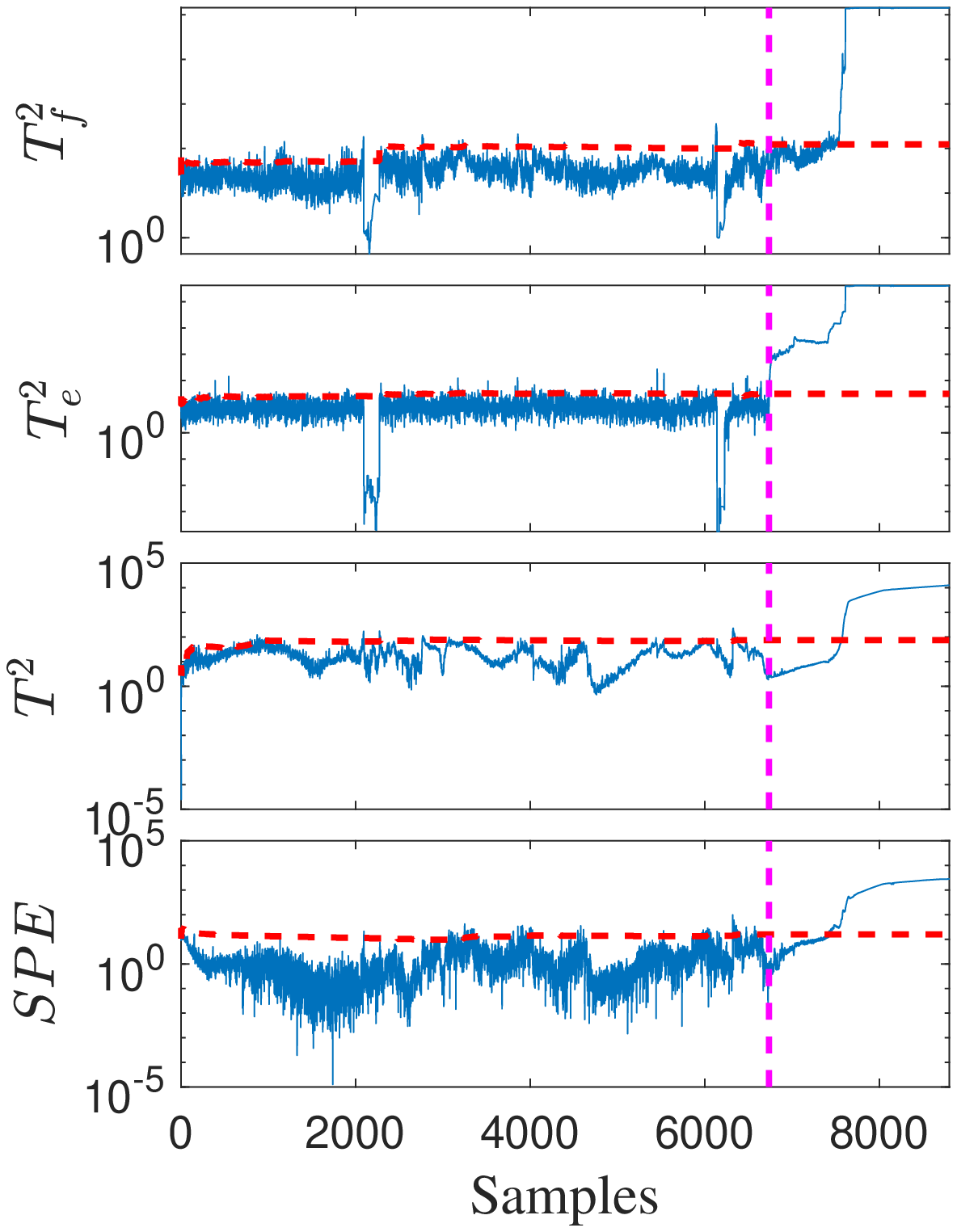}}}
    	\subfigure{\label{fault1-3}}\addtocounter{subfigure}{-1}
    	{\subfigure[RCA-RPCA-EWC]{\includegraphics[width=0.312\textwidth]{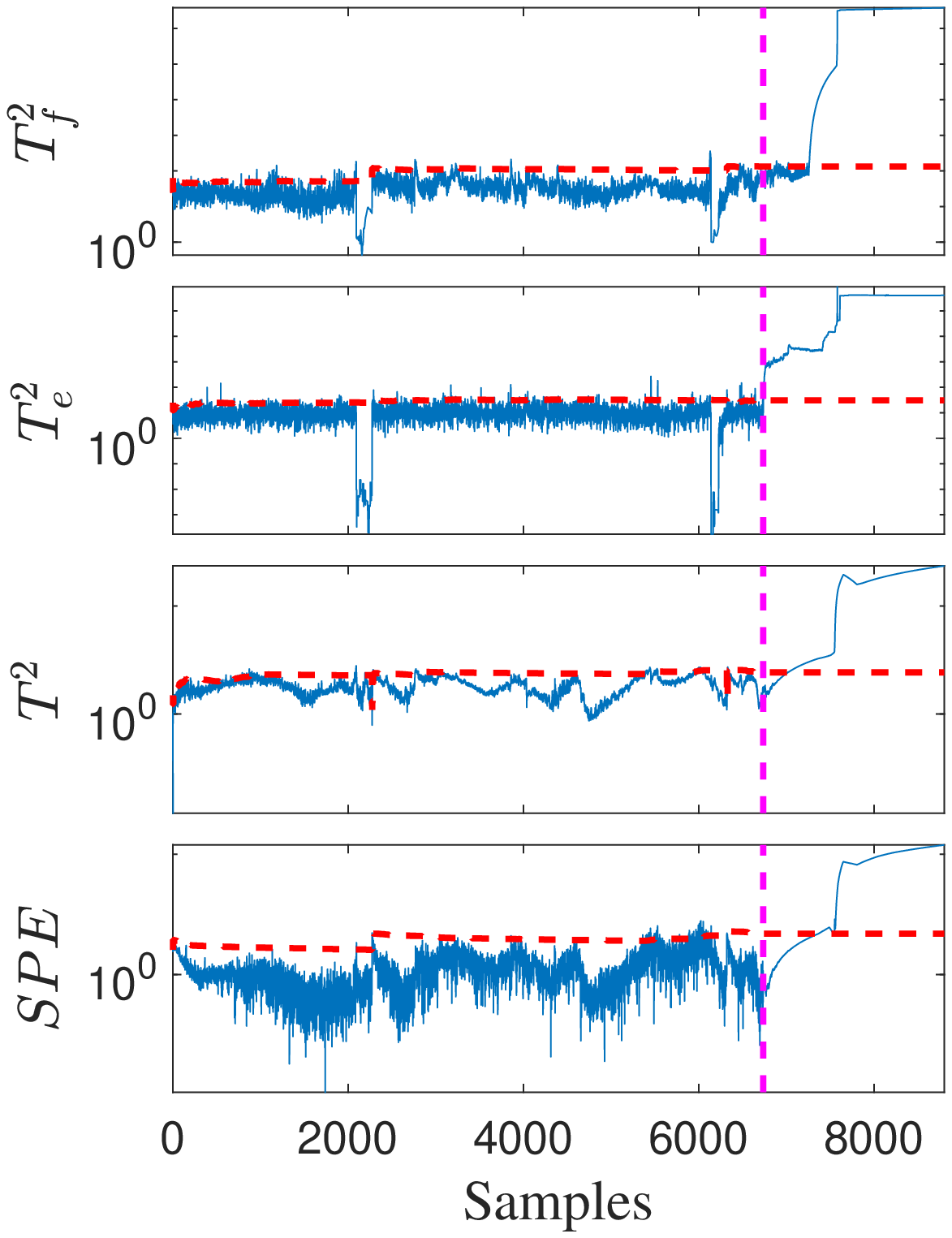}}}
    \caption{Monitoring charts of Case 1} \label{case1}
    \end{figure*}

   \begin{figure*}[!htbp]
    	\centering
    	\subfigure{\label{fault2-1}}\addtocounter{subfigure}{-1}
    {\subfigure[Recursive CA]{\includegraphics[width=0.312\textwidth]{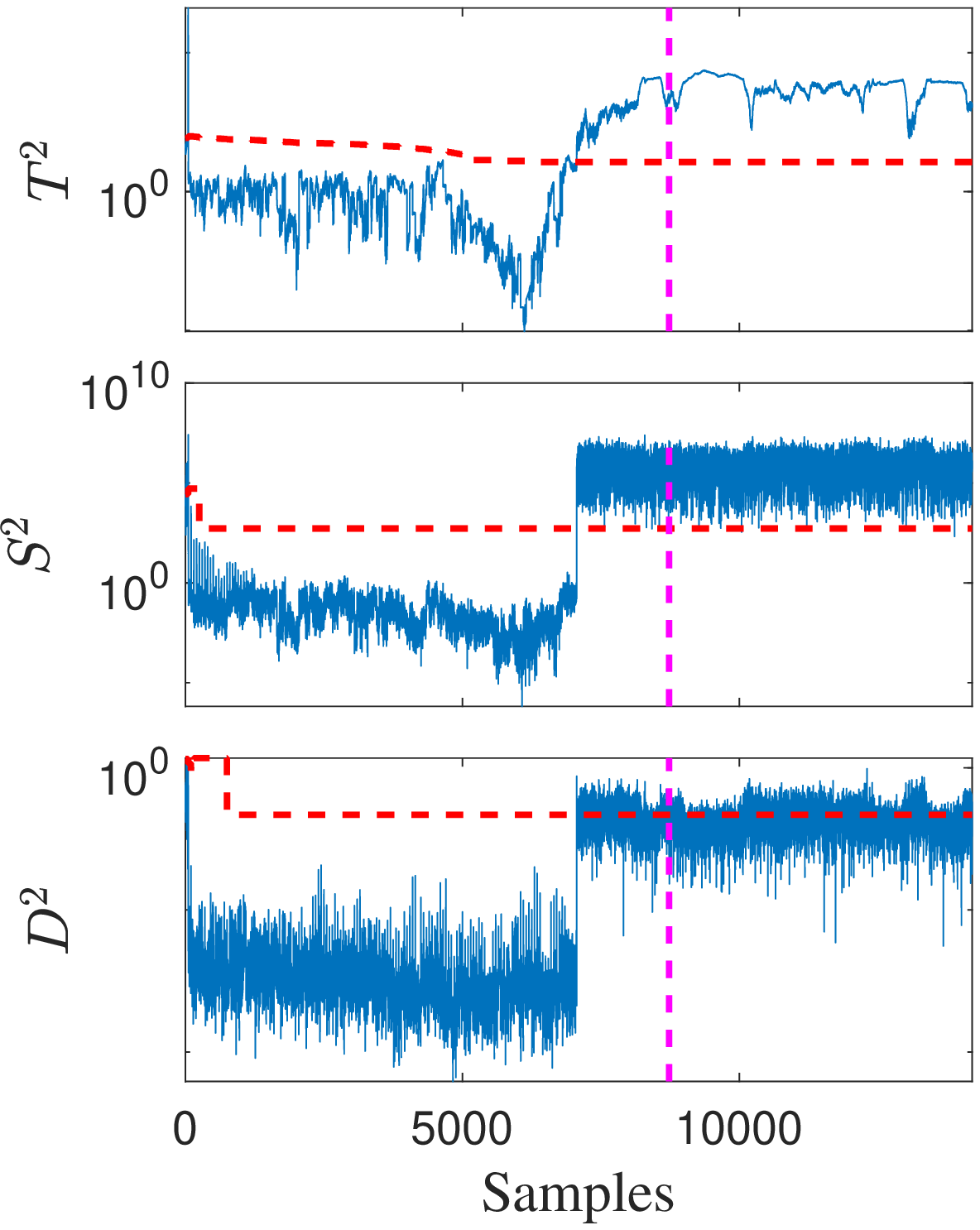}}}
    	\subfigure{\label{fault2-2}}\addtocounter{subfigure}{-1}
    	{\subfigure[RCA-RPCA]{\includegraphics[width=0.320\textwidth]{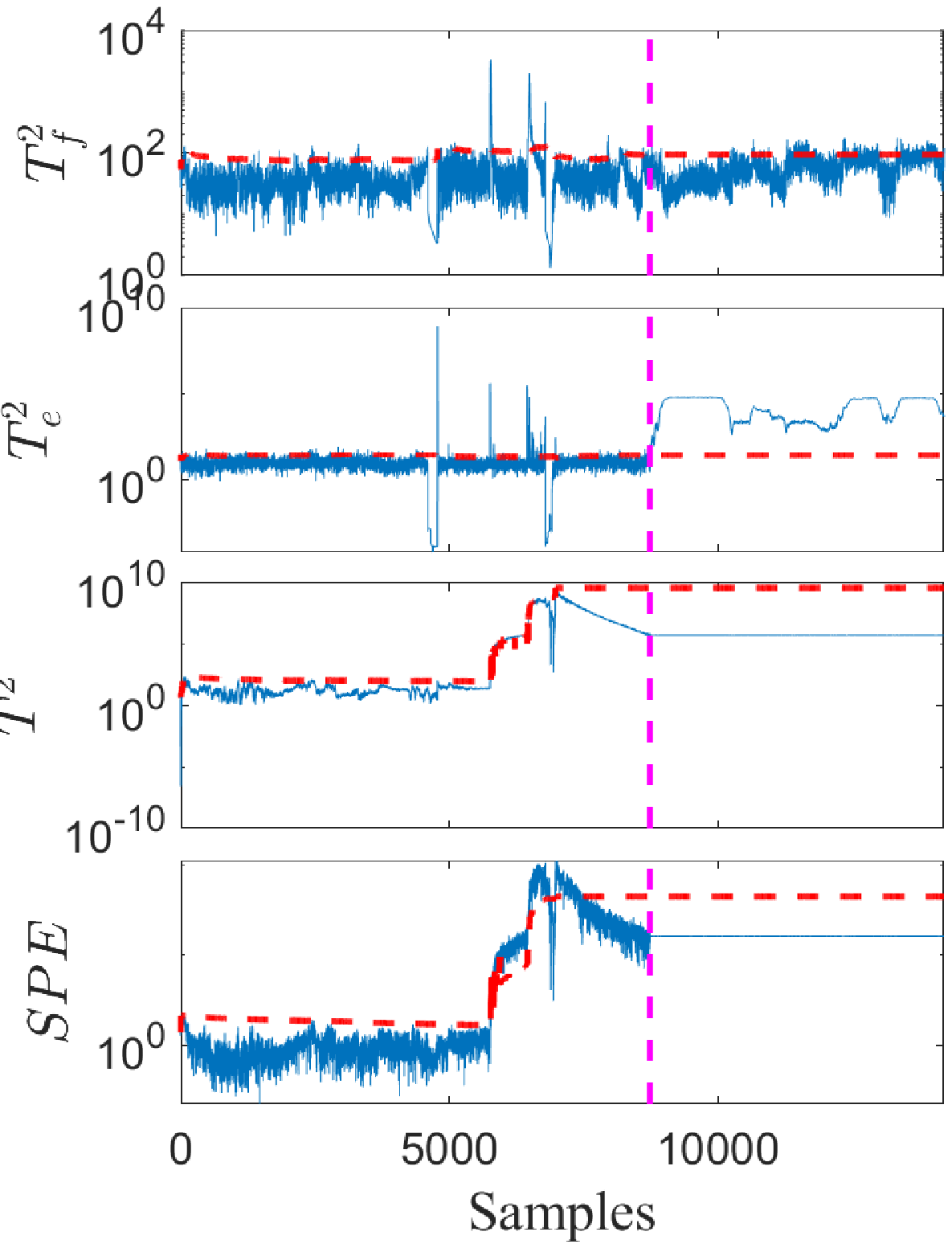}}}
    	\subfigure{\label{fault2-3}}\addtocounter{subfigure}{-1}
    	{\subfigure[RCA-RPCA-EWC]{\includegraphics[width=0.32\textwidth]{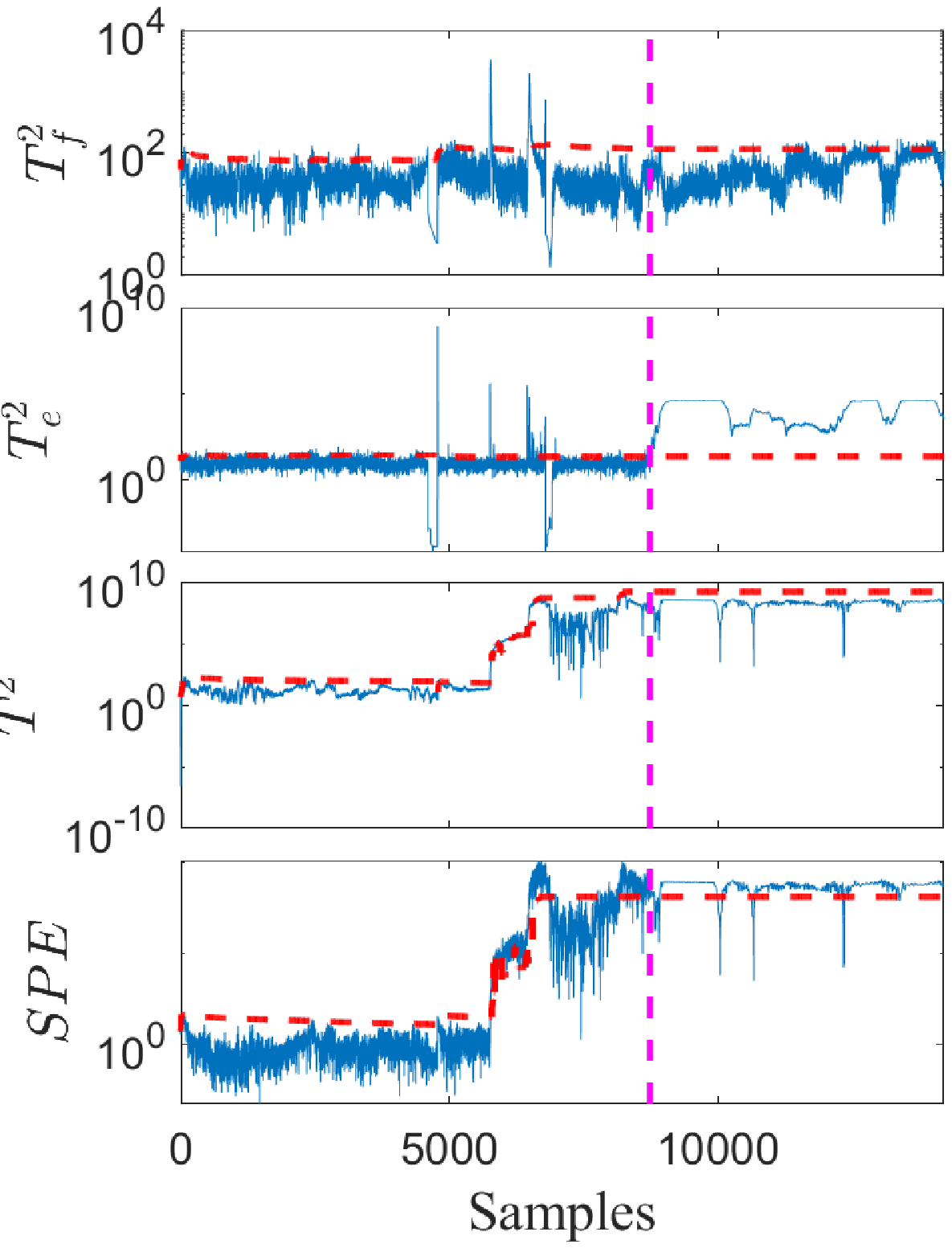}}}
    \caption{Monitoring charts of Case 2} \label{case2}
    \end{figure*}

   \begin{figure*}[!htbp]
    	\centering
    	\subfigure{\label{fault3-1}}\addtocounter{subfigure}{-1}
    	{\subfigure[Recursive CA]{\includegraphics[width=0.312\textwidth]{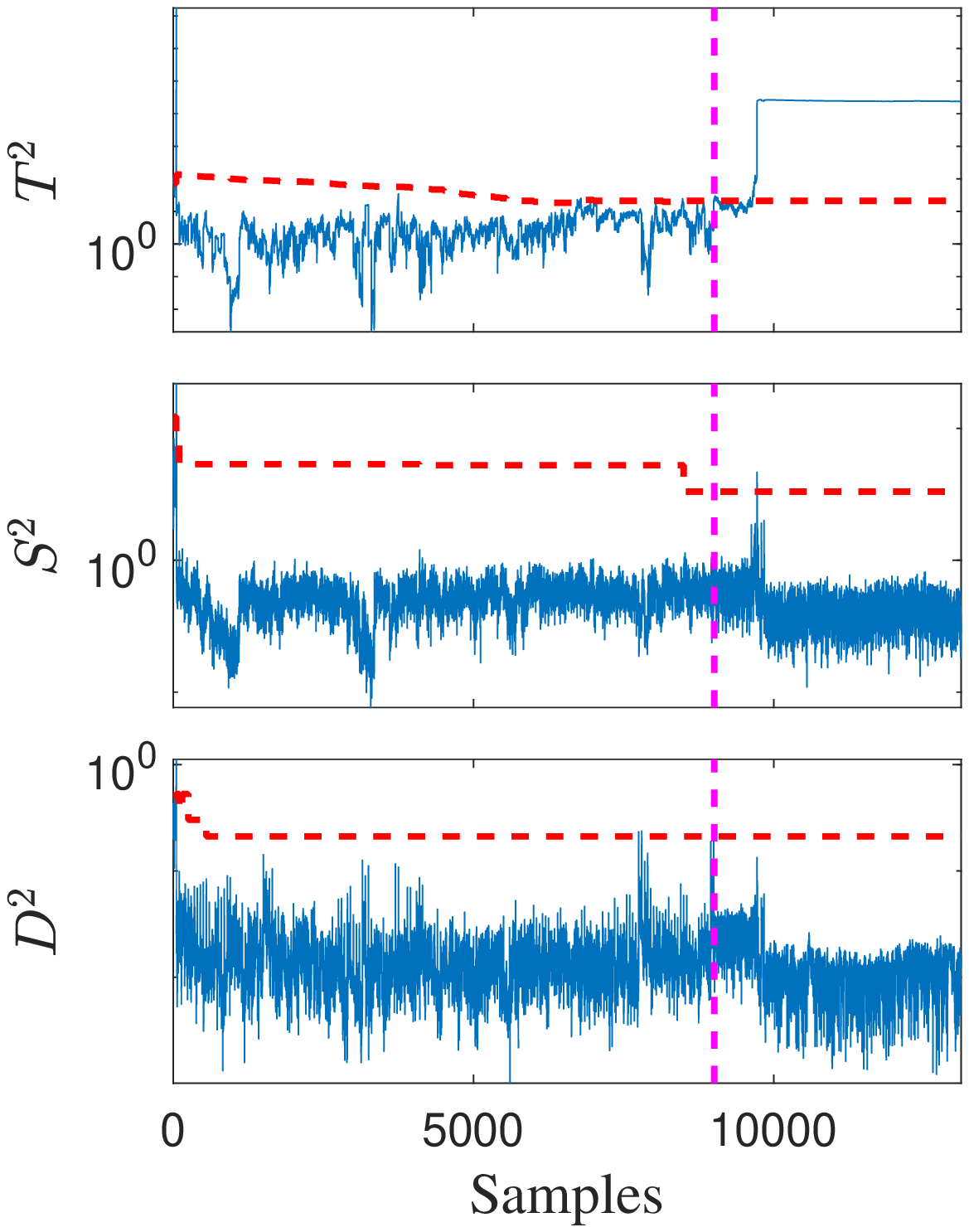}}}
    	\subfigure{\label{fault3-2}}\addtocounter{subfigure}{-1}
    	{\subfigure[RCA-RPCA]{\includegraphics[width=0.3250\textwidth]{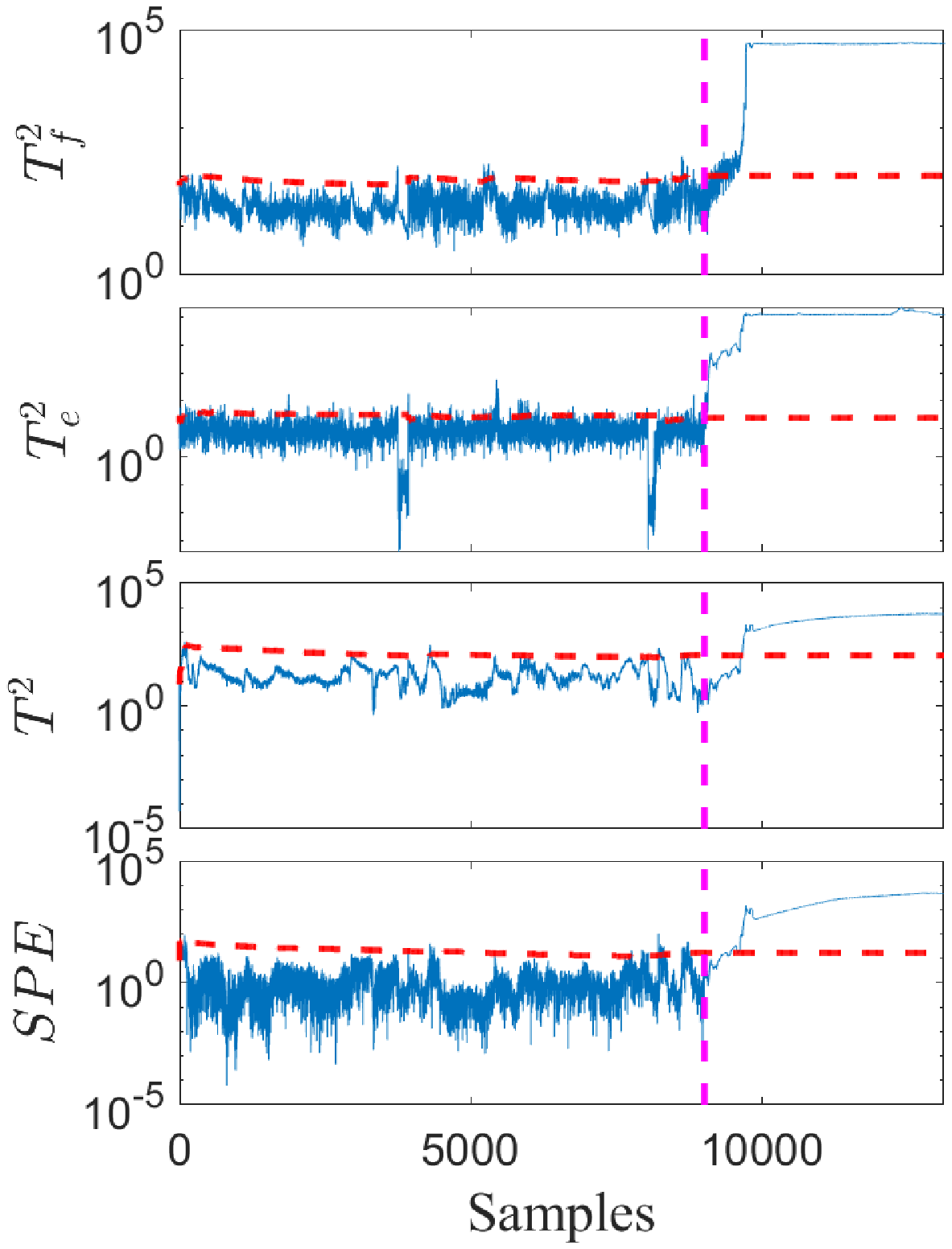}}}
    	\subfigure{\label{fault3-3}}\addtocounter{subfigure}{-1}
    	{\subfigure[RCA-RPCA-EWC]{\includegraphics[width=0.325\textwidth]{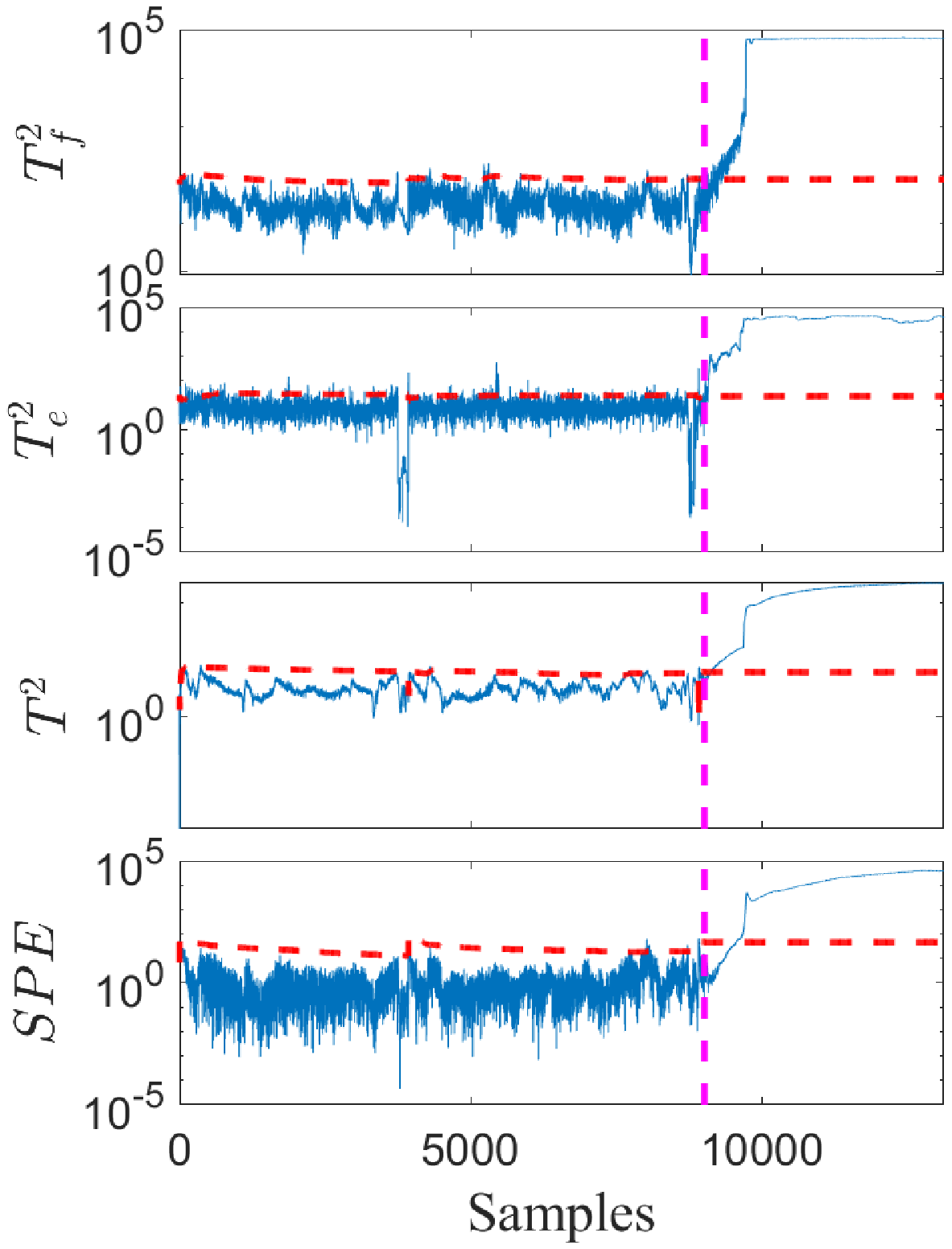}}}
    \caption{Monitoring charts of Case 3} \label{case3}
    \end{figure*}

  \subsection{Simulation analysis}

In this paper, we compare recursive CA \cite{yu2020recursive} with the proposed RCA to illustrate the virtues of real-time update. Then, the proposed RCA-RPCA is compared with RCA-RPCA-EWC to illustrate the superiorities of EWC. Note that RCA-RPCA and RCA-RPCA-EWC share the same RCA algorithm proposed in Section \ref{section:recursiveCA}.

Three indicators are considered to evaluate the performance, namely, fault detection rates (FDRs), false alarm rates (FARs) and detection delay (DD). The calculation method can refer to \cite{zhang2019an}. DD refers to the number of samples that the fault is detected later than the recorded fault time, which is valuable and significant for practical industrial systems. The monitoring consequences of  three case are described in Figs. \ref{case1}-\ref{case3}, respectively. Note that the pink vertical line represents the practical fault time instant.

The monitoring charts of Case 1 are presented  in Fig. \ref{case1}.
Recursive CA \cite{yu2020recursive} fails to detect the fault accurately and the FDR is $59.31\%$. Besides, $S^2$ and $D^2$  can not distinguish novelty from normal dynamic changes in Fig. \ref{fault1-1}. For the proposed RCA, $T_e^2$ can detect the fault precisely and timely, and the detection delay is about 1 minute. In the time period, where $T^2_e$ and $T^2_f$ change significantly, the type of coal changes and the current cointegration relationship may be broken. Thus, the CA model needs to be retrained from scratch based on the newly collected data. During this time period, the process is monitored by the current CA model, and new data are collected to build the initial CA model that is appropriate for the new material. Thus, $T^2_e$ and $T^2_f$ recover to be stable quickly. Compared with RPCA, RPCA-EWC provides better performance in Fig. \ref{fault1-3} and  the FDR of $T^2$ is $86.31\%$. However, the FDR of RPCA is $66.81\%$, $19.5\%$ lower than RPCA-EWC.  According to $T^2_e$ and $T^2_f$, it is observed that the type of coal varies and a new model is built before the fault occurs, which indicates that RPCA is not trained  enough and can not track the system change actually.

For Case 2, the monitoring results are depicted in Fig. \ref{case2}.  For the proposed RCA method, $T_e^2$ can detect the fault accurately and the FDR is $99.98\%$.  In Figs. \ref{fault2-2} and \ref{fault2-3},  $T_e^2$ and $T_f^2$ change sharply twice. According to the coal records and original data analysis, the first sudden change of two statistics originates from the switch of coal type, while the second abrupt change is attributed to the critical parameters adjusted artificially. Compared with RPCA, SPE of RPCA-EWC enables to detect the fault precisely and the FDR is $93.86\%$. The short-term dynamic of two types of coal has a certain degree of similarity, and the significant information of previous coal is preserved and beneficial to monitor other coal. The FARs of SPE are relatively high because RPCA is not able to track the rapid system change at the initial stage. The system is judged as normal because SPE returns to normal quickly. Regardless of the false alarms caused by this situation, the FARs of SPE are $5.20\%$ and $7.62\%$ in Figs. \ref{fault2-2}-\ref{fault2-3}, respectively. However, recursive CA \cite{yu2020recursive} misidentifies the normal parameter variations as anomalies in Fig. \ref{fault2-1} and the FAR is up to $20\%$. It is insensitive to faults that are orthogonal to cointegration space and only dynamic information is not enough to monitor the process effectively.

For Case 3, the monitoring consequences are exhibited in Fig. \ref{case3}. The recursive CA \cite{yu2020recursive} detects the fault inexactly and the FDR of $T^2$ is $86.21\%$. The FDRs of $S^2$ and $D^2$ are 0, and thus it is meaningless to mention delay detection. The proposed RCA can detect the fault accurately and the FDRs are more than $99\%$ in Figs. \ref{fault3-2}-\ref{fault3-3}. For RPCA-EWC in Fig. \ref{fault3-3}, the FDR of $T^2$ is $97.20\%$, which indicates that the significant information of the previous coal is preserved by EWC and  beneficial to deliver excellent monitoring performance. However, the FDR of RPCA is less than $90\%$ in Fig. \ref{fault3-2}.

The evaluation indexes of three cases are summarized in Table \ref{Table1}. Compared with recursive CA \cite{yu2020recursive}, the proposed RCA is more sensitive to normal changes from human intervention and raw materials changing. This phenomenon occurs owing to several factors: a) The variables are selected and divided based on prior knowledge and ADF test, which is more universal and accurate than just ADF; b) Critical stationary variables, which are sensitive to raw materials changing, are utilized to establish the $T^2_f$ statistic; c) In \cite{yu2020recursive}, the model is updated based on a block of data and the monitoring performance is effected by the block length,  
while the proposed RCA model is updated in real time and more compatible with the current operating system.  In addition, compared with RCA-RPCA, RCA-RPCA-EWC preserves significant information of previous influential parameters and avoids dramatic performance degradation when similar operating modes revisit.

\section{Conclusion}\label{section:conclusion}
In this paper,  RCA-RPCA-EWC was developed to monitor the general nonstationary processes, where the proposed RCA is updated in real time and able to distinguish the real faults from normal system deviations. To avoid potential ill-conditioning issue of matrix inversion, several calculation techniques are adopted and the RCA issue is settled with low computational burden. As RCA is insensitive to faults that are orthogonal to cointegration space, the remaining information of RCA together with other short-term dynamic information is monitored by RPCA to establish a comprehensive monitoring framework. When the system enters a new operating mode, EWC is employed to strengthen the significant information of previous operating modes and avoid the abrupt performance degradation for future similar operating modes. Besides, the test statistics are constructed based on RCA and the prior knowledge, which are sensitive to mode identification.  Compared with recursive CA \cite{yu2020recursive} and RCA-RPCA, the effectiveness and superiorities of the proposed method are illustrated by a practical industrial pulverizing system.


In future, we will investigate the quality-related nonstationary process monitoring. Besides, graceful forgetting will be considered as forgetting older modes is essential to make space for learning newer modes.

\appendix

\subsection{RPCA for process monitoring}\label{rank1-FOEP}

In this paper, RPCA is implemented based on on rank-1 modification with first-order perturbation (FOP). Detailed information can refer to \cite{elshenawy2010efficient}.

At $k+1$ instant, the sample $\boldsymbol x_{k+1}^0 \in \mathbb{R}^{m_2}$ is collected. Then, the mean $ \boldsymbol \mu$ and standard deviation $\sigma_1,\cdots,\sigma_{m_2}$ are updated as:
\begin{equation}\label{meank}
 \boldsymbol \mu_{k+1} = \alpha_{k+1} \boldsymbol \mu_k +(1-\alpha_{k+1})\boldsymbol x_{k+1}^0
\end{equation}
 \begin{equation}\label{stdk}
\sigma_{i,k+1}^2 = \alpha_{k+1} \sigma_{i,k+1}^2 + (1-\alpha_{k+1})(x_{i,k+1}^0-\mu_{i,k+1})^2
\end{equation}
where $i = 1,\cdots, m_2$, $\alpha_{k+1} =  \frac{k}{{k + 1}}$ is the forgetting factor. The sample is normalized  as
\begin{equation}
{\boldsymbol x}_{k+1} = (\boldsymbol x_{k+1}^0- \boldsymbol \mu_{k+1}) \boldsymbol \Sigma_{k+1}^{-1}
\end{equation}
where $\boldsymbol \Sigma_{k+1} = diag(\sigma_{1,k+1},\cdots,\sigma_{m_2,k+1})$.

Based on rank-1 modification and FOP, the eigenvectors and eigenvalues are updated as:
\begin{equation}\label{RECUR1}
  \boldsymbol P_{k+1}=\boldsymbol P_{k}\left( \boldsymbol I+\boldsymbol Q_V \right)
\end{equation}
\begin{equation}\label{RECUR2}
\boldsymbol \Lambda_{k+1}=\alpha_{k+1} \boldsymbol \Lambda _{k}+\left( 1-\alpha_{k+1} \right) \boldsymbol Q_{\Lambda}
\end{equation}
Define the rank-1 matrix $\boldsymbol \kappa _{k+1}=\boldsymbol P_{k}^{T}\boldsymbol x_{k+1}$, the diagonal matrix $\boldsymbol Q_{\varLambda}$ and $\boldsymbol Q_V$ are calculated by:
\begin{equation}\label{rpca_qlambda}
  Q_{\varLambda}\left( i,i \right) =\kappa_i
\end{equation}
\begin{equation}\label{rpca_qv}
  \begin{cases}
	Q_V\left( i,j \right) =\frac{\kappa_i \kappa_j}{\tau_j+\kappa_{j}^{2}-\tau_i+\kappa_{i}^{2}},i\ne j\\
	Q_V\left( i,i \right) =0\\
\end{cases}
\end{equation}
where $\kappa_i$ is the $i$th element of $\boldsymbol \kappa_{k+1}$, $\tau_i$ and $\tau_j$ are the $i$th and $j$th corresponding elements of $k \boldsymbol \varLambda_{k}$.

\subsection{Recursive computation of $\varDelta \boldsymbol A$ and  $\varDelta \boldsymbol B$}\label{section:deltaAB}

We illustrate the computation of $ \varDelta \boldsymbol A_{k+1} $ and $ \varDelta \boldsymbol B_{k+1} $. Take the component  $\boldsymbol D_{k+1}^{T} \boldsymbol E_{1,k} $ of  $ \varDelta \boldsymbol A_{1,k+1} $ as an example, we show that the computation of  $ \varDelta \boldsymbol A_{1,k+1} $ is $O(m_1^2)$, which is independent of the number of existing samples.
\begin{equation}\label{dek}
\boldsymbol D_{k+1}^{T} \boldsymbol E_{1,k}=\boldsymbol d_{k+1}^{T}\varDelta \boldsymbol x_{k+1}^{p}\boldsymbol J_{k}^{T}\boldsymbol E_{1,k}
\end{equation}
 If we compute (\ref{dek}) directly, the computation is  $O(km_1^2)$ and increases linearly with the emerging samples. We need to get the recursive form of $ \boldsymbol J_{k}^{T}\boldsymbol E_{1,k} $.
 \begin{equation}\label{jek}
\begin{aligned}
 & \boldsymbol J_{k}^{T} \boldsymbol E_{1,k}\\
=&\left[ \begin{array}{c}
\boldsymbol J_{k-1} \boldsymbol {\tilde{J}}_k\\
\varDelta \boldsymbol x_{k}^{p} \boldsymbol R_{k-1} \boldsymbol {\tilde{J}}_k\\
\end{array} \right] ^T\left[ \begin{array}{c}
\boldsymbol E_{1,k-1}- \boldsymbol J_{k-1}\left( \varDelta \boldsymbol x_{k}^{p} \right) ^T \boldsymbol h_k\\
\boldsymbol h_k\\
\end{array} \right]
\\
=&\boldsymbol {\tilde{J}}_{k}^{T} \boldsymbol J_{k-1}^{T}\left(\boldsymbol E_{1,k-1}-\boldsymbol J_{k-1}\left( \varDelta \boldsymbol x_{k}^{p} \right) ^T \boldsymbol h_k \right) +\varDelta \boldsymbol x_{k}^{p}\boldsymbol R_{k-1} \boldsymbol {\tilde{J}}_k \boldsymbol h_k
\\
=&\boldsymbol {\tilde{J}}_{k}^{T} \boldsymbol  J_{k-1}^{T} \boldsymbol  E_{1,k-1}-\boldsymbol {\tilde{J}}_{k}^{T}\boldsymbol J_{k-1}^{T} \boldsymbol  J_{k-1}\left( \varDelta \boldsymbol x_{k}^{p} \right) ^T \boldsymbol h_k
\end{aligned}
 \end{equation}

 The recursion of $ \boldsymbol  J_{k}^{T} \boldsymbol  J_{k} $ is:
 \begin{equation}\label{jjk}
 \boldsymbol J_{k}^{T}  \boldsymbol J_k=\boldsymbol {\tilde{J}}_{k}^{T} \boldsymbol J_{k-1}^{T} \boldsymbol J_{k-1}\boldsymbol {\tilde{J}}_k+\boldsymbol {\tilde{J}}_{k}^{T} \boldsymbol R_{k-1}^{T}\left( \varDelta \boldsymbol x_{k}^{p} \right) ^T\varDelta \boldsymbol x_{k}^{p} \boldsymbol R_{k-1} \boldsymbol {\tilde{J}}_k
 \end{equation}
Based on (\ref{dek}-\ref{jjk}), $ \boldsymbol D_{k+1}^{T} \boldsymbol E_{1,k} $ can be calculated recursively and the computational complexity is $ O(m_1^2) $ by reasonable arrangement of matrix calculation order. It is obviously true because each component contains at least one vector. Similarly, other components of $ \varDelta \boldsymbol A_{k+1} $ and  $ \varDelta  \boldsymbol B_{k+1} $ need  $ O(m_1^2) $. In conclusion, the computational complexity of $ \varDelta \boldsymbol A_{k+1} $ and  $ \varDelta \boldsymbol B_{k+1} $ is $ O(m_1^2) $.

\subsection{The recursive calculation of $\boldsymbol K$}\label{section:recursion_K}
$\boldsymbol{K}_{k+1}$ is calculated by:
\begin{equation}
\boldsymbol{K}_{k+1}=\boldsymbol{B}_{k+1}^{-\frac{1}{2}}\,\,=\,\,\left( {\alpha }_{k+1}\boldsymbol{B}_{k}+\left( 1-{\alpha }_{k+1} \right) \boldsymbol{\varDelta B}_{k+1} \right) ^{-\frac{1}{2}}
\end{equation}
Considering that  $\boldsymbol B$ is a block diagonal matrix with $\boldsymbol B = \left[ {\begin{array}{*{20}{c}}
{{\boldsymbol B_1}}&\boldsymbol 0\\
\boldsymbol 0&{{\boldsymbol B_2}}
\end{array}} \right]$, we mainly introduce the recursion of $ \boldsymbol B_1 $, and $ \boldsymbol B_2 $  can be updated similarly.
\begin{equation}\label{invB1}
\begin{aligned}
  &\boldsymbol{K}_{1,k+1} =\boldsymbol{B}_{1,{k}+1}^{-\frac{1}{2}}\,\, \\
=&\,\,\left( \alpha _{{k}+1}\boldsymbol{B}_{1,{k}}+\left( 1-\alpha _{{k}+1} \right) {\varDelta \boldsymbol B}_{1,{k}+1} \right) ^{-\frac{1}{2}}
\\
=&\left( \left( \sqrt{\alpha _{k+1}}\boldsymbol B_{1,k}^{\frac{1}{2}} \right) \left( \boldsymbol I+\frac{1-\alpha _{k+1}}{\alpha _{k+1}}\boldsymbol B_{1,k}^{-\frac{1}{2}}{\varDelta \boldsymbol B}_{1,{k}+1} \right. \right.\\
  &\left. \left. \left(\boldsymbol B_{1,k}^{-\frac{1}{2}}\right)^T \right) \left( \sqrt{\alpha _{k+1}}\boldsymbol B_{1,k}^{\frac{1}{2}} \right)^T \right) ^{-\frac{1}{2}}
\\
=&\frac{1}{\sqrt{\alpha _{k+1}}}\left( \boldsymbol I+\frac{1-\alpha _{k+1}}{\alpha _{k+1}}\boldsymbol B_{1,k}^{-\frac{1}{2}}{\varDelta \boldsymbol B}_{1,{k}+1} \boldsymbol B_{1,k}^{-\frac{1}{2}} \right) ^{-\frac{1}{2}}\boldsymbol B_{1,k}^{-\frac{1}{2}}\\
=&\frac{1}{\sqrt{\alpha _{k+1}}}\left( \boldsymbol I+\frac{1-\alpha _{k+1}}{\alpha _{k+1}}\boldsymbol K_{1,k}{\varDelta \boldsymbol B}_{1,{k}+1} \boldsymbol K^T_{1,k} \right) ^{-\frac{1}{2}}\boldsymbol K_{1,k}
\\
=&\frac{1}{\sqrt{\alpha _{k+1}}}\boldsymbol {\tilde K}_{1,k+1}^{-\frac{1}{2}}\boldsymbol K_{1,k}
\end{aligned}
\end{equation}
where ${\boldsymbol {\tilde K}_{1,k+1}} = \boldsymbol I+\frac{1-\alpha _{k+1}}{\alpha _{k+1}}\boldsymbol K_{1,k}{\varDelta \boldsymbol B}_{1,{k}+1} \boldsymbol K^T_{1,k}$.
The key is to calculate $\boldsymbol {\tilde K}_{1,k+1}^{-\frac{1}{2}}$.
${\varDelta \boldsymbol B}_{1,{k}+1}$ is symmetric and the rank is no more than 2. Thus, it can be reformulated into
\begin{equation}
\begin{aligned}
&\varDelta \boldsymbol  B_{1,k+1} \\
=&\left[ \begin{matrix}
\boldsymbol q_{1,k+1}&		 \boldsymbol q_{2,k+1}\\
\end{matrix} \right] \left[ \begin{matrix}
\beta _{1,k+1}&		\\
&		\beta _{2,k+1}\\
\end{matrix} \right] \left[ \begin{matrix}
\boldsymbol q_{1,k+1}&		 \boldsymbol q_{2,k+1}\\
\end{matrix} \right] ^T\\
=&\boldsymbol Q_{1,k+1} \boldsymbol \varXi_{1,k+1} \boldsymbol Q_{1,k+1}^T
\end{aligned}
\end{equation}
where $\beta _{1,k+1}$ and $ \beta _{2,k+1} $ are non-zero eigenvalues if $ rank({\varDelta \boldsymbol B}_{1,{k}+1}) =2 $. $\boldsymbol q_{1,k+1} $ and $\boldsymbol q_{2,k+1} $ are the corresponding eigenvectors. If the rank is $ 1 $, then $\beta _{1,k+1}$ or $ \beta _{2,k+1} $ is $ 0 $. Thus,
\begin{equation} \label{trans1}
\begin{aligned}
 &{\boldsymbol {\tilde K}_{1,k+1}}\\
= & \boldsymbol I+\frac{1-\alpha _{k+1}}{\alpha _{k+1}} \boldsymbol K_{1,k}\boldsymbol Q_{1,k+1} \boldsymbol \varXi_{1,k+1} \boldsymbol Q_{1,k+1}^T \boldsymbol K^T_{1,k}\\
=&  \boldsymbol I+\frac{1-\alpha _{k+1}}{\alpha _{k+1}} \left(\boldsymbol K_{1,k} \boldsymbol Q_{1,k+1}\right)\boldsymbol \varXi_{1,k+1} \left(\boldsymbol K_{1,k} \boldsymbol Q_{1,k+1} \right)^T\\
=& \boldsymbol I + {\tilde{\boldsymbol Q}}_{1,k+1} {\tilde{\boldsymbol \varXi}}_{1,k+1} {\tilde{ \boldsymbol Q}}_{1,k+1}^T
\end{aligned}
\end{equation}
where $ {\tilde{\boldsymbol Q}}_{1,k+1} = \boldsymbol K_{1,k} \boldsymbol Q_{1,k+1} \in \mathbb{R}^{m \times 2} $, $ {\tilde{\boldsymbol \varXi}}_{1,k+1} = \frac{1-\alpha _{k+1}}{\alpha _{k+1}} \boldsymbol \varXi_{1,k+1}$ and  $ rank({\tilde{\boldsymbol \varXi}}_{1,k+1} ) \leqslant 2$.
As  $ \boldsymbol B_{1,k+1} $ is obviously positive definite, then $ \boldsymbol I + {\tilde{\boldsymbol Q}}_{1,k+1} {\tilde{\boldsymbol \varXi}}_{1,k+1} {\tilde{ \boldsymbol Q}}_{1,k+1}^T $ is also positive definite. For convenience, let $\tilde{\boldsymbol Q}_{1,k+1}=\left[ \begin{matrix}
\tilde{\boldsymbol q}_{1,k+1}&		\tilde{\boldsymbol q}_{2,k+1}\\
\end{matrix} \right]  $,  $ {\tilde{\boldsymbol \varXi}}_{1,k+1} =\left[ \begin{matrix}
\tilde{\beta}_{1,k+1}&		\\
&		\tilde{\beta}_{2,k+1}\\
\end{matrix} \right]  $.
 We select the calculation manner of  $\boldsymbol {\tilde K}_{1,k+1}^{-\frac{1}{2}}$
 based on the rank of $ {\tilde{\boldsymbol \varXi}}_{1,k+1}$.

1) $ rank({\tilde{\boldsymbol \varXi}}_{1,k+1} ) = 1$.

Let $ \beta _{1,k+1} \neq 0  $ and $ \beta _{2,k+1} =0 $, here
${\tilde{\boldsymbol \varXi}}_{1,k+1} =  \tilde {\beta}_{1,k+1}$, $ {\tilde{\boldsymbol Q}}_{1,k+1} =  \tilde{\boldsymbol q}_{1,k+1} $.
Then,
\begin{equation}
\begin{aligned}
&\left( \boldsymbol I + {\tilde{\boldsymbol Q}}_{1,k+1} {\tilde{\boldsymbol \varXi}}_{1,k+1} {\tilde{ \boldsymbol Q}}_{1,k+1}^T \right)^{-\frac{1}{2}}\\
=& \left( \boldsymbol I +   \tilde{\beta} _{1,k+1} \tilde{\boldsymbol q}_{1,k+1} \tilde{\boldsymbol q}_{1,k+1}^T  \right)^{-\frac{1}{2}} \\
=&\boldsymbol I+\frac{\tilde{\boldsymbol q}_{1,k+1}\tilde{\boldsymbol q}_{1,k+1}^{T}}{\tilde{\boldsymbol q}_{1,k+1}^{T}\tilde{\boldsymbol q}_{1,k+1}}\left( \frac{1}{\sqrt{1+\tilde{\beta}_{1,k+1}\tilde{\boldsymbol q}_{1,k+1}^{T}\tilde{\boldsymbol q}_{1,k+1}}}-1 \right) \\
=&\boldsymbol  I+ \gamma_{1,k+1} \tilde{\boldsymbol q}_{1,k+1}\tilde{\boldsymbol q}_{1,k+1}^{T}
\end{aligned}
\end{equation}
where $ \gamma_{1,k+1} = \frac{1}{\tilde{\boldsymbol q}_{1,k+1}^{T}\tilde{\boldsymbol q}_{1,k+1}}\left( \frac{1}{\sqrt{1+\tilde{\beta}_{1,k+1}\tilde{\boldsymbol q}_{1,k+1}^{T}\tilde{\boldsymbol q}_{1,k+1}}}-1 \right)$.
Thus, the recursion of $ \boldsymbol K_1 $ is
\begin{equation}\label{inv5B_rank1}
\boldsymbol{K}_{1,{k}+1}=  \frac{1}{\sqrt{\alpha _{k+1}}}  \left( \boldsymbol  I+ \gamma_{1,k+1} \tilde{\boldsymbol q}_{1,k+1}\tilde{\boldsymbol q}_{1,k+1}^{T}  \right)
\boldsymbol K_{1,k}
\end{equation}

 2) $ rank({\tilde{\boldsymbol \varXi}}_{1,k+1} ) = 2$.

  The formula (\ref{trans1}) can be further reformulated into
 \begin{equation}
 \begin{aligned}
 &\boldsymbol I + {\tilde{\boldsymbol Q}}_{1,k+1} {\tilde{\boldsymbol \varXi}}_{k+1} {\tilde{ \boldsymbol Q}}_{1,k+1}^T\\
= & \boldsymbol I +\gamma_{1,k+1} \tilde{\boldsymbol q}_{1,k+1}\tilde{\boldsymbol q}_{1,k+1}^{T} +\gamma_{2,k+1} \tilde{\boldsymbol q}_{2,k+1}\tilde{\boldsymbol q}_{2,k+1}^{T} \\
=& \check{\boldsymbol Q}_{1,k+1} \boldsymbol{\check{\varLambda}}_{1,k+1}\check{\boldsymbol Q}_{1,k+1}^{T}
 \end{aligned}
 \end{equation}
 where $  \check{\boldsymbol Q}_{1,k+1} $ is the eigen matrix with $ \check{\boldsymbol Q}_{1,k+1}^{T}\check{\boldsymbol Q}_{1,k+1}=\boldsymbol I $, $ \boldsymbol{\check{\varLambda}}_{1,k+1} $ contains the eigenvalues.
(\ref{trans1}) is realized by two successive rank-1 modification with FOP \cite{elshenawy2010efficient}. Thus,
(\ref{invB1}) is further calculated:
\begin{equation}
\boldsymbol K_{1,k+1}= \frac{1}{\sqrt{\alpha _{k+1}}}\boldsymbol{\check{\varLambda}}^{-\frac{1}{2}}_{1,k+1} {\check{\boldsymbol Q}}_{1,k+1}^{T} \boldsymbol K_{1,k}
\end{equation}

$\boldsymbol K_{2,k+1}$ can be calculated in the similar way. Thus, $\boldsymbol K_{k+1} =\left[ \begin{matrix}
\boldsymbol K_{1,k+1}&		\\
&		\boldsymbol K_{2,k+1}\\
\end{matrix} \right]  $.

\subsection{Solution of RPCA-EWC}\label{solution_RPCAEWC}
The formula (\ref{total_loss}) can be reformulated as
 \begin{equation}\label{objective_original}
 	\begin{aligned}
 	\mathcal{J}(\boldsymbol {P}) 	=& tr(\boldsymbol  P^T {\boldsymbol \Omega} \boldsymbol P)-tr(\boldsymbol  P^T  \boldsymbol X_2^T \boldsymbol X_2 \boldsymbol  P) -2  tr(\boldsymbol  P^T {\boldsymbol \Omega} \boldsymbol  P^*_0) \\
 		& + \underbrace{\{tr(\boldsymbol X_2^T \boldsymbol X_2)+ tr({\boldsymbol  P^*_0}^T {\boldsymbol \Omega} \boldsymbol  P^*_0)\}}_{constant}
 	\end{aligned}
 \end{equation}

Let $G(\boldsymbol P) =  tr(\boldsymbol  P^T {\boldsymbol \Omega} \boldsymbol P)-2 tr(\boldsymbol  P^T {\boldsymbol \Omega} \boldsymbol  P^*_0)$, $H(\boldsymbol P) = tr(\boldsymbol  P^T  \boldsymbol X_2^T \boldsymbol X_2 \boldsymbol  P)$. Thus, $\mathcal{J}(\boldsymbol {P}) = G(\boldsymbol P)-H(\boldsymbol P)+constant $.
The minimization of (\ref{total_loss}) is equivalent to
 \begin{equation}\label{objective_function}
 \begin{aligned}
& \mathop{min}\limits_{\boldsymbol P} \quad G(\boldsymbol P)-H(\boldsymbol P) \\
& s.t. \qquad \boldsymbol  P^T \boldsymbol  P = \boldsymbol I \in R^{l \times l}
 \end{aligned}
 \end{equation}
Since $G(\boldsymbol P)$ and $H(\boldsymbol P)$ are convex,  the objective function (\ref{objective_function}) is actually DC programming problem \cite{Souza2015Global,Voorhis2003Difference}. DC programming includes linearizing the convex function and solving the convex function as follows \cite{zhang2020multimode}.  

 Assume that $\boldsymbol P_i$ is the solution at $i$th iteration, we approximate the second part $H(\boldsymbol P)$ by linearizing
 \begin{equation}
H_l(\boldsymbol P) = H(\boldsymbol P_i)+\langle \boldsymbol P-\boldsymbol P_i, \boldsymbol U_i\rangle
 \end{equation}
Since the  subgradient $\boldsymbol U \in \partial H(\boldsymbol P) = 2 \boldsymbol X_2^T \boldsymbol X_2 \boldsymbol P$, let $\boldsymbol U_i = 2\boldsymbol X_2^T \boldsymbol X_2 \boldsymbol P_i$. Then, (\ref{objective_function}) is approximated by
 \begin{equation}\label{objective_subproblem}
\boldsymbol P_{i+1} \doteq \underset{\boldsymbol P^T \boldsymbol P = \boldsymbol I}{\arg \min} \quad G(\boldsymbol P)-<\boldsymbol P,\boldsymbol U_i>
 \end{equation}
Since ${\boldsymbol \Omega}$ is semidefinite, let ${\boldsymbol \Omega} = \boldsymbol L^T \boldsymbol L$ and $\boldsymbol L$ is the triangle matrix \cite{zhang2020multimode}. Thus, we can get
  \begin{equation}\label{formulated_solution}
 \begin{aligned}
   & G(\boldsymbol P)-<\boldsymbol P,\boldsymbol U_i> \\
  =& tr(\boldsymbol P^T {\boldsymbol \Omega} \boldsymbol P)-2tr(\boldsymbol P^T {\boldsymbol \Omega} \boldsymbol  P_0^*)-2 tr(\boldsymbol P^T \boldsymbol X_2^T \boldsymbol X_2 \boldsymbol P_i)\\
  = &  {\langle \boldsymbol L\boldsymbol P, \boldsymbol L\boldsymbol P\rangle}-2 \langle \boldsymbol L\boldsymbol P, \boldsymbol L\boldsymbol  P_0^*+ (\boldsymbol L^T)^{-1}\boldsymbol X_2^T \boldsymbol X_2 \boldsymbol P_i  \rangle \\
  =& \lVert  \boldsymbol Z_i- \boldsymbol L\boldsymbol P\rVert _F^2- \lVert  \boldsymbol Z_i\rVert _F^2
 \end{aligned}
 \end{equation}
 where $ \boldsymbol Z_i = \boldsymbol L\boldsymbol  P_0^*+ (\boldsymbol L^T)^{-1}\boldsymbol X_2^T \boldsymbol X_2 \boldsymbol P_i $ is constant at $i+1$th iteration \cite{zhang2020multimode}.

Then, (\ref{formulated_solution})  is equivalent to \cite{huang2014robust,zhang2020multimode}
\begin{equation}\label{final_objective2}
\boldsymbol P_{i+1} = \underset{\boldsymbol P^T \boldsymbol P = \boldsymbol I}{\arg \min} \quad \lVert  \boldsymbol P-\boldsymbol L^T \boldsymbol Z_i\rVert _F^2
\end{equation}
 Let $\boldsymbol Y_i=\boldsymbol L^T \boldsymbol Z_i= {\boldsymbol \Omega} \boldsymbol  P_0^*+ \boldsymbol X_2^T \boldsymbol X_2 \boldsymbol P_i$. According to the lemma in \cite{huang2014robust}, we can obtain that $ \boldsymbol P_{i+1} = \boldsymbol W_i \boldsymbol I_{m,l}\boldsymbol V_i^T $, where $\boldsymbol W_i \in \mathbb R^{m \times m}$ and $\boldsymbol V_i \in \mathbb R^{l \times l} $ are  left and right singular values of the singular vector decomposition of $\boldsymbol Y_i$ \cite{zhang2020multimode}. The procedure is summarized in Algorithm \ref{EWC_solution}.


\bibliography{my_references}
\bibliographystyle{ieeetr}

\end{document}